\documentclass{aastex6}
\usepackage{mkfig}

\begin{document}

\title{A {\em Chandra}/LETGS Survey of Main Sequence Stars}

\author{Brian E. Wood\altaffilmark{1}, J. Martin Laming\altaffilmark{1},
  Harry P. Warren\altaffilmark{1}, Katja Poppenhaeger\altaffilmark{2}}

\altaffiltext{1}{Naval Research Laboratory, Space Science Division,
  Washington, DC 20375, USA; brian.wood@nrl.navy.mil}
\altaffiltext{2}{Astrophysics Research Centre, School of Mathematics and
  Physics, Queen's University Belfast, Belfast BT7 1NN, UK}


\begin{abstract}

     We analyze the X-ray spectra of 19 main sequence
stars observed by {\em Chandra} using its LETGS configuration.
Emission measure (EM) distributions are computed based on emission
line measurements, an analysis that also yields evaluations of coronal
abundances.  The use of newer atomic physics data results in
significant changes compared to past published analyses.  The stellar
EM distributions correlate with surface X-ray flux ($F_X$) in a
predictable way, regardless of spectral type.  Thus, we provide EM
distributions as a function of $F_X$, which can be used to estimate
the EM distribution of any main sequence star with a measured
broadband X-ray luminosity.  Comparisons are made with solar EM
distributions, both full-disk distributions and spatially resolved
ones from active regions (ARs), flares, and the quiet Sun.  For
moderately active stars, the slopes and magnitudes of the EM
distributions are in excellent agreement with those of solar ARs for
$\log T<6.6$, suggesting that such stars have surfaces completely
filled with solar-like ARs.  A stellar surface covered with solar
X-class flares yields a reasonable approximation for the EM
distributions of the most active stars.  Unlike the EM distributions,
coronal abundances are very spectral-type dependent, and we provide
relations with surface temperature for both relative and absolute
abundances.  Finally, the coronal abundances of the exoplanet host
star $\tau$~Boo~A (F7~V) are anomalous, and we propose that this is
due to the presence of the exoplanet.


\end{abstract}


\section{Introduction}

     Cool main sequence stars are universally observed to be
surrounded by hot coronae ($T=10^{6-7}$ K), which represent
the outermost atmospheric layers of such stars.  The nature and
origin of the surprisingly hot coronae has been a focal point of
solar/stellar atmosphere research for decades.  Stellar coronae are
best studied at short wavelengths, as most emission from hot coronal
material comes out in X-ray and extreme ultraviolet (EUV) radiation.
Most of what we know about stellar coronae comes from soft X-ray
observations from a series of spacecraft dating back to the 1970s
\citep{mg04a,mg09}.  This legacy continues
today with the {\em Chandra} and {\em XMM-Newton} spacecraft, which
have both been observing the X-ray sky since 1999.

     Only so much can be learned about coronae from simple broadband
flux measurements.  More detailed study requires high resolution
X-ray spectroscopy.  Both {\em Chandra} and XMM carry gratings that
provide spectra of unprecedented quality.  The Reflection
Grating Spectrometers (RGS) on XMM observe from $5-35$~\AA.
{\em Chandra} carries two gratings, the High Energy Transmission
Grating (HETG) and the Low Energy Transmission Grating (LETG).  The
former is generally paired with the ACIS-S detector to make the HETG
Spectrometer (HETGS), which observes from $1.2-31$~\AA.  The latter is
generally paired with the HRC-S detector to make the LETG Spectrometer
(LETGS), observing from $5-175$~\AA.  With its particularly broad
wavelength range, {\em Chandra}/LETGS is capable of observing
significantly more lines of more atomic species than either
{\em Chandra}/HETGS or XMM/RGS.  Thus, we will here be focusing on
the analysis of {\em Chandra}/LETGS data.

     Analysis of emission lines observed in a coronal X-ray spectrum
involves the reconstruction of the coronal emission measure (EM)
distribution.  This process yields two crucial diagnostics.  The
first is the EM distribution itself, which describes the distribution of
temperature in the corona, and the second involves the measurement of
element abundances in the corona, which can be different from
photospheric abundances.  Both the temperature and abundance
diagnostics are crucial for studying the mechanism(s) behind
coronal heating.

     As for the abundance diagnostic, the solar corona and wind exhibit
an abundance pattern where elements with low first ionization
potential (FIP) have abundances that are enhanced relative to elements
with high FIP \citep{rvs95,uf00,jml15}.
Similar ``FIP effects'' have been observed for some
stellar coronae \citep{jjd97,jml99}, but in
other cases coronal abundances appear to be close to photospheric
\citep{jjd95,ma01,jsf09}.
Finally, there are a number of cases where an ``inverse FIP effect''
is observed, where low-FIP elements are {\em depleted} relative to
high-FIP elements \citep{mg01,dph01,ma03,jr05}.  On the Sun, an inverse
FIP effect has recently been detected for the first time near sunspots
during flares \citep{gad15}.

     Significant effort has been expended to study how coronal abundances
vary with activity level and spectral type.  Initial attention focused
on the connection between high activity and inverse FIP, as
notoriously active M dwarfs and active binaries tend to have inverse
FIP \citep{cl08}.  In a survey of early G dwarfs, \citet{at05}
found inverse FIP or no FIP effect for the youngest
and most active stars with ages less than 300~Myr, but solar-like
FIP effects at older ages.  The importance of spectral type for
coronal abundances becomes clear if attention is focused only on main
sequence stars, particularly if extremely active stars with X-ray
luminosities (in ergs~s$^{-1}$) of $\log L_X>29$ are ignored.  For such
stars, a surprisingly tight relation between FIP bias and spectral
type is found, with M dwarfs having an inverse
FIP effect, which reduces toward no FIP effect at a mid-K spectral type,
and then drifts toward a solar-like FIP effect for early G dwarfs
\citep{bew10,bew12}.  We will refer to this
relation as the ``FIP-Bias/Spectral-Type'' (FBST) relation.  This
relation implies that {\em all} M dwarfs have inverse FIP, not just
the very active ones.  Thus, the vast majority of main sequence stars in
the Galaxy (all but the most active) are presumed to follow the
FBST relation.

     The element fractionation that occurs in the process of coronal
heating is potentially a crucial diagnostic of this heating, and
any successful coronal heating model should be able to explain
the fractionation patterns that are observed on the Sun and other
stars.  Currently the only theoretical framework for explaining both
a FIP effect and an inverse FIP effect involves the presence of
ponderomotive forces induced by Alfv\'{e}n waves passing through
coronal loops, with the direction of the force along the loop
depending on where the waves are introduced and where they reflect
within the loops \citep{jml04,jml09,jml12,bew13}.
If correct, this model would strongly support an important role
for Alfv\'{e}n waves (or their generation) in coronal heating.

     Returning to the temperature diagnostic characterized by the
EM distributions themselves, coronal temperature is an even more
direct diagnostic of coronal heating than the abundances, with higher
temperatures implying more intense heating.  Observations clearly
show that more active stars, with higher X-ray luminosities,
systematically have higher coronal temperatures, implying that
increases in stellar activity are not simply a matter of filling
the stellar surface with more and more identical active regions
\citep{mg97,js97,at05}.  The most recent empirical analysis is that of
\citet{cpj15}, who find an impressively tight relation
between mean coronal temperature and X-ray surface flux for cool main
sequence stars of all types:  $T_{cor}=0.11 F_X^{0.26}$, with $T_{cor}$
in MK units and $F_X$ in ergs~cm$^{-2}$~s$^{-1}$.

     Such analyses rely on reducing the coronal temperature
distribution to just one or two temperatures.  In reality, however,
coronal temperature distributions seem to be more or less
continuous, as opposed to singly or doubly valued.  With
nearly 20 years of data now acquired by {\em Chandra} and XMM, it
should now be possible to use detailed EM distributions to more
precisely describe how coronal temperature changes with increasing
activity, thereby providing more detailed constraints for coronal
heating models.  For example, one interpretation of the increase in
$T_{cor}$ with activity is that this is indicative of the increasing
dominance of flare-like emission as stellar activity increases.
Another interpretation is that it is indicative of the increasing
prevalence of a population of hotter and presumably larger coronal
loops \citep[e.g.,][]{mg04a,fr14}.  A precise assessment of how
the EM distribution evolves as activity increases could be helpful for
distinguishing between such interpretations, and this is a central
goal of our project.

     To accomplish this goal, we conduct a survey of
all main sequence stars observed with {\em Chandra}/LETGS.
We measure EM distributions using uniform analysis procedures, and
assuming consistent atomic data.  We also measure coronal
abundances for our sample of stars, and study how they vary with
activity and spectral type.  This work overlaps strongly with
work already done on the FBST relation (see above), but we
expand this in a number of important ways.  For example, the FBST
relation defined above involves measurements of {\em relative} coronal
abundances, particularly abundances of high-FIP elements relative to
the best measured low-FIP element, Fe.  However, we here also
assess whether {\em absolute} abundances, for Fe in particular, vary
with spectral type like the relative abundances.

    A final motivation for this project is that our
sample of {\em Chandra}/LETGS spectra includes two particularly
noteworthy recent observations obtained by us.
One is an observation of $\eta$~Lep (F1~V) in 2017~December, which
is the earliest type main sequence star successfully observed by
{\em Chandra} with grating spectroscopy.  This helps us to extend
the FBST relation to earlier spectral types, and test whether stars
with very thin convection zones have significantly different EM
distributions.  Also, in 2017 February-March {\em Chandra}/LETGS
observed the exoplanet host star $\tau$~Boo~A (F7~V).  This observation
allows us to assess whether a very close-in, massive exoplanet can
affect coronal temperatures and abundances in a way that makes the
star clearly anomalous in our survey.

\section{Sample Definition and Data Reduction}

\begin{table}[t]
\scriptsize
\begin{center}
\caption{Main Sequence Stars Observed with {\em Chandra}/LETGS}
\begin{tabular}{lccccccccccccc} \hline \hline
Star & Spectral & Dist. & Radius    & $T_{eff}$ & $\log L_X$  &
  \multicolumn{8}{c}{Photospheric Abundances (Relative to Solar)$^a$} \\
\cline{7-14}
     &   Type   &   (pc)   &(R$_{\odot}$)&  (K)    &(erg s$^{-1}$)&
   C & O & Mg & Si & Ca & Fe & Ni & Ref. \\
\hline
$\eta$ Lep   & F1 V & 14.9 & 1.56 & 6902 & 28.50 &
  -0.07 & -0.01 &  0.11 & -0.17 & -0.16 & -0.17 & -0.14 & 1 \\
$\pi^3$ Ori  & F6 V & 8.07 & 1.32 & 6424 & 28.99 &
   0.19 &  0.16 & -0.02 &  0.06 &  0.10 &  0.00 & -0.05 & 2 \\
$\tau$ Boo A & F7 V & 15.6 & 1.47 & 6387 & 28.76 &
   0.33 &  0.30 & (0.33)&  0.33 & (0.33)&  0.33 &  0.22 & 3 \\
$\pi^1$ UMa  & G1 V & 14.4 & 0.97 & 5768 & 28.99 &
  -0.11 & -0.04 & -0.26 & -0.09 & -0.11 & -0.25 & -0.24 & 2 \\
EK Dra       &G1.5 V& 35.8 & 0.93 & 5845 & 30.06 &
  (0.00)& (0.00)& (0.00)& (0.00)& (0.00)& (0.00)& (0.00)& ... \\
$\alpha$ Cen A&G2 V & 1.34 & 1.22 & 5792 & 26.99 &
   0.18 &  0.25 &  0.39 &  0.32 &  0.40 &  0.12 &  0.20 & 2 \\
$\xi$ Boo A  & G8 V & 6.70 & 0.86 & 5570 & 28.91 &
  -0.10 & -0.09 & -0.26 & -0.10 & -0.35 & -0.26 & -0.24 & 2 \\
70 Oph A     & K0 V & 5.09 & 0.83 & 5202 & 28.09 &
  -0.10 &  0.03 &  0.09 &  0.18 &  0.09 & -0.05 &  0.06 & 2 \\
AB Dor A     & K0 V & 15.2 & 0.79 & 5047 & 30.06 &
  (0.00)& (0.00)& (0.00)& (0.00)& (0.00)& (0.00)& (0.00)& ... \\
$\alpha$~Cen~B&K1 V & 1.34 & 0.86 & 5231 & 27.32 &
   0.28 &  0.37 &  0.40 &  0.46 &  0.47 &  0.27 &  0.40 & 2 \\
36 Oph A     & K1 V & 5.99 & 0.69 & 5192 & 28.02 &
  -0.40 & -0.14 & -0.28 & -0.07 & -0.15 & -0.30 & -0.20 & 2 \\
36 Oph B     & K1 V & 5.99 & 0.59 & 5192 & 27.89 &
  -0.40 & -0.14 & -0.28 & -0.07 & -0.15 & -0.30 & -0.20 & 2 \\
$\epsilon$~Eri&K2 V & 3.22 & 0.74 & 5076 & 28.31 &
  -0.24 & -0.04 & -0.03 & -0.01 & -0.01 & -0.06 & -0.06 & 2 \\
$\xi$ Boo B  & K4 V & 6.70 & 0.61 & 4620 & 28.08 &
  -0.10 & -0.09 & -0.26 & -0.10 & -0.35 & -0.26 & -0.24 & 2 \\
61 Cyg A     & K5 V & 3.49 & 0.67 & 4374 & 27.03 &
 (-0.33)&(-0.33)& -0.18 & -0.29 & -0.36 & -0.33 & -0.39 & 4 \\
70 Oph B     & K5 V & 5.09 & 0.67 & 4450 & 27.97 &
  -0.10 &  0.03 &  0.09 &  0.18 &  0.09 & -0.05 &  0.06 & 2 \\
61 Cyg B     & K7 V & 3.49 & 0.60 & 4044 & 26.97 &
 (-0.38)&(-0.38)& -0.06 & -0.33 & -0.40 & -0.38 & -0.43 & 4 \\
AU Mic       & M1 Ve& 9.91 & 0.61 & 3684 & 29.36 &
  (0.00)& (0.00)& (0.00)& (0.00)& (0.00)& (0.00)& (0.00)& ... \\
AD Leo       &M4.5 Ve&4.89 & 0.38 & 3336 & 28.70 &
  (0.00)& (0.00)& (0.00)& (0.00)& (0.00)& (0.00)& (0.00)& ... \\
\hline
\end{tabular}
\end{center}
\tablerefs{(1) \citet{ky11}; (2) \citet{cap04};
  (3) \citet{gg07}; (4) \citet{pj15}.}
\tablecomments{$^a$Values in parentheses assumed rather than measured.}
\normalsize
\end{table}
     Our data sample is defined by existing {\em Chandra}/LETGS
spectra of main sequence stars deemed to be of sufficient quality for
our purposes.  Table~1 lists the 19 targets with spectra that have
numerous enough detectable emission lines for an EM analysis to be
performed.  The targets are listed in order of spectral type, ranging
from $\eta$~Lep (F1~V) to AD~Leo (M4.5~Ve).  The stars cover a wide
range of activity levels, with X-ray luminosities
(column 6 in Table~1) ranging from $\log L_X=26.99$ (in erg~s$^{-1}$)
for $\alpha$~Cen~A to $\log L_X=30.06$ for EK~Dra and AB~Dor~A.  These
X-ray luminosities are for the canonical {\em ROSAT} PSPC soft X-ray
bandpass of $0.1-2.4$~keV (e.g., $5-120$~\AA), and are measured directly 
from the LETGS spectra themselves.

     The fourth column of Table~1 lists radii for our stars, which are
necessary to compute X-ray surface fluxes ($F_X$).  Recent work provides
evidence in favor of $F_X$ being a preferable measure of stellar activity
compared with $L_X$ or $L_X/L_{bol}$ \citep{cpj15,rsb17}.  Our default source
of radius estimation is the relation of \citet{tgb78}, but many
radii are taken from more direct measurements
\citep{pk03,jm08,jdg11,tsb12a,tsb12b}.
The fifth column lists photospheric effective
temperatures ($T_{eff}$).  Most of these are from published spectral
analyses of the individual stars \citep{jav05,jh09,uh15}.
The remainder are estimated using the $T_{eff}$ versus $B-V$ relation
of \citet{jav05}, or for stars later than early-K taken from
Table~5 in \citet{mjp13} or the $T_{eff}$ versus $V-K$
relation of \citet{awm15}.

     Our measured coronal abundances will have to be compared with
photospheric abundances, so Table~1 lists photospheric abundance
measurements for our stars.  Following the usual convention, the
abundances are listed logarithmically relative to solar photospheric
abundances, so a value of 0.0
corresponds to an abundance equal to that
of the solar photosphere.  Throughout the paper our default
solar photospheric reference abundances will be those of
\citet{ma09}.  For three companion stars (36~Oph~B,
$\xi$~Boo~B, and 70~Oph~B) we assume the companion has the same
abundances as the primary.  No photospheric abundances are available
for the two M dwarfs in the sample (AU~Mic and AD~Leo), as the
formation of molecules makes photospheric abundance measurements
very difficult.  We are forced to simply assume solar photospheric
abundances for those stars.  Likewise, photospheric abundance
measurements are also very difficult for the two rapidly rotating
stars EK~Dra and AB~Dor~A, due to line blending induced by rotational
broadening.  We once again simply assume solar abundances
for those stars, consistent with crude estimates from optical spectra
\citep{spj07,ov87}.  Finally,
for 61~Cyg~AB there are no measurements for C and N, so we simply
assume values identical to Fe.

\begin{table}[t]
\scriptsize
\begin{center}
\caption{{\em Chandra}/LETGS Observations}
\begin{tabular}{lccc} \hline \hline
Star & Obs. ID & Start Time & Exp. Time \\
     &         &            &   (ksec) \\
\hline
$\eta$ Lep      & 20130 & 2017 Dec 15 21:12:58 & 116.3 \\
                & 20884 & 2017 Dec 11 17:03:23 &  38.0 \\
$\pi^3$ Ori     & 12324 & 2010 Nov 9 12:17:00  &  57.2 \\
                & 13184 & 2010 Nov 21 19:23:36 &  19.9 \\
$\tau$ Boo AB   & 17715 & 2017 Feb 27 22:41:08 &  47.9 \\
                & 20019 & 2017 Mar 4 23:43:24  &  28.5 \\
                & 20020 & 2017 Mar 5 17:41:14  &  15.0 \\
$\pi^1$ UMa     & 23    & 2000 Jan 15 6:14:22  &  30.0 \\
EK Dra          & 1884  & 2001 Mar 19 3:20:38  &  65.5 \\
$\alpha$ Cen AB & 29    & 1999 Dec 24 10:38:20 &  79.6 \\
                & 7432  & 2007 Jun 4 7:15:44   & 117.1 \\
                & 12332 & 2011 Jun 8 12:32:19  &  78.5 \\
$\xi$ Boo AB    & 8899  & 2008 May 2 11:20:48  &  93.3 \\
70 Oph AB       & 4482  & 2004 Jul 19 22:59:01 &  77.9 \\
AB Dor AB       & 3762  & 2002 Dec 10 21:25:18 &  85.3 \\
36 Oph AB       & 4483  & 2004 Jun 1 10:22:33  &  77.4 \\
$\epsilon$ Eri  & 1869  & 2001 Mar 21 7:17:12  & 105.3 \\
61 Cyg AB       & 13651 & 2012 Feb 13 20:19:33 & 187.7 \\
AU Mic          & 8894  & 2008 Jun 26 12:08:31 &  49.4 \\
AD Leo          & 975   & 2000 Oct 24 15:06:16 &  48.1 \\
\hline
\end{tabular}
\end{center}
\normalsize
\end{table}
     Table~2 lists the individual {\em Chandra}/LETGS observations that
we have to work with, and we now describe the data reduction procedures
used to process the data.  Rather than use the default processed spectrum,
we process the data ourselves, using version 4.9 of the CIAO software
provided by the {\em Chandra} X-ray Center (CXC) \citep{af06}.  Many of our
targets are binary stars with two separate resolved sources, and a tailored
data processing is necessary in such cases anyway.  An LETGS observation
consists of a zeroth-order image of the target, with plus and minus
order X-ray spectra dispersed in opposite directions from the image.
The zeroth-order image is used to establish the
central reference point for a spectral extraction, and it is also useful
for providing a broadband X-ray light curve for the observation, allowing
flares to be identified.  Although we do find some modest flares within
the data, in all cases the flares are too weak or brief
to contribute greatly to the counts of the overall integrated spectrum,
so no attempt is made to remove any of the flares in the spectral
extraction procedure.

     For an isolated star, our default procedure assumes a conservatively
broad extraction window of $\Delta s=30$ pixels, increasing to
$\Delta s=90$ for wavelengths greater than 90~\AA.  The increase in window
size is necessary to account for the worsening spatial resolution at
higher wavelengths farther from the aim point.  Broad
background windows are extracted on both sides of the source spectrum
in order to estimate the background level, which is rather high
for the HRC-S detector.  Even though the average background can be
accurately measured and subtracted from the source spectrum,
the background is a significant source of noise.  The HRC-S detector has
limited energy resolution, but some amount of pulse height filtering is
possible and can be used to reduce the background.  We use the
recommended background filter, following the relevant CIAO analysis
thread on the CXC
website\footnote{http://cxc.cfa.harvard.edu/ciao/threads/spectra\_letghrcs}.
Still, for some of the fainter
targets with noisier spectra, we find it necessary to decrease $\Delta s$
to minimize the background noise further.  We use extraction windows as
narrow as $\Delta s=12$ pixels in some cases for this reason, although
we always expand to $\Delta s=90$ for wavelengths greater than 90~\AA.
The downside of narrow windows is a degradation of photometric accuracy,
but for noisy spectra this cost is more than balanced by the improvement in
signal-to-noise (S/N) due to the decreased background.

     For binary stars it is necessary to avoid overlapping
spectral extraction windows, so $\Delta s$ can be limited by the stellar
separation for close binaries.  In such cases, above 90~\AA\ it is
generally possible
to expand the extraction window in only one direction to avoid
overlapping windows.  There will be some degree of unresolved source
blending at these higher wavelengths.  We extract separate spectra for the
two components of all binaries identified in Table~2 (e.g., the stars
with ``AB'' in their names).  In most cases, both components are
considered in our target sample (see Table~1), with two exceptions,
$\tau$~Boo~B (M2~V) and AB~Dor~B (M5~V+M5-6~V).  For $\tau$~Boo~B and
AB~Dor~B, we deem their spectra to be too noisy with too few detected
lines for us to include these sources in our target list.

     After background subtraction, the final step is to coadd the plus
and minus orders.  In many cases, it is necessary to shift either the
plus or minus order spectra by up to four pixels before coaddition
in order to line up the emission lines.  This is an indication of the
uncertainties in the LETGS wavelength calibration, which can vary
in an unpredictable manner from observation to observation.

     There are multiple observations listed in Table~2 for four stars.
For $\eta$~Lep, $\pi^3$~Ori, and $\tau$~Boo we simply coadd the
observations to create a final spectrum.  However, for
$\alpha$~Cen~AB the three observations are taken far apart in time,
with the two stars at different points in their activity cycles.
The $\alpha$~Cen system has been monitored regularly in X-rays by
both {\em Chandra} and XMM, and using these data \citet{tra14} estimates
activity cycle periods of $19.2\pm 0.7$ and $8.1\pm 0.2$~yr for
$\alpha$~Cen~A and B, respectively.  For $\alpha$~Cen~A, the
first observation (ID \#29) occurred near an activity cycle maximum
in 1998, while the last two occurred closer to a minimum in 2008.  We
therefore consider two separate $\alpha$~Cen~A spectra in our analysis,
an $\alpha$~Cen~A(hi) spectrum associated with observation ID \#29, and
an $\alpha$~Cen~A(lo) spectrum that is a coaddition of the other
two observations (ID's \#7432 and \#12332).  Similarly, for
$\alpha$~Cen~B it is the third observation that is near an activity
cycle maximum, with the other two near minima.  Thus, we construct
an $\alpha$~Cen~B(hi) spectrum from observation ID \#12332, and an
$\alpha$~Cen~B(lo) spectrum that is a coaddition of the other two
observations (ID's \#29 and \#7432).

\section{Line Identification and Measurement}

     The first step in our analysis of the {\em Chandra}/LETGS spectra
is to identify and measure emission lines in the spectra.  Our approach
is to focus first on the highest quality spectra in our sample to establish
the largest possible list of clearly detected and identified lines for
the EM analysis, and then we search for only these lines in the
other spectra.  This is in effect a Bayesian approach to finding
lines in the noisier spectra, as we will only be looking at the precise
wavelengths where the better quality data have informed us that we
might reasonably find a line, which in turn allows us to be less
conservative about claiming at least a marginal detection.  A $1\sigma$
flux excess at exactly the right wavelength where a line is expected
is far more likely to be a detection than a $1\sigma$
flux excess at some random location.  Even for
nondetections, we estimate upper limits for line fluxes, which will
be considered in the EM analysis described below.

\begin{figure}[t]
\plotfiddle{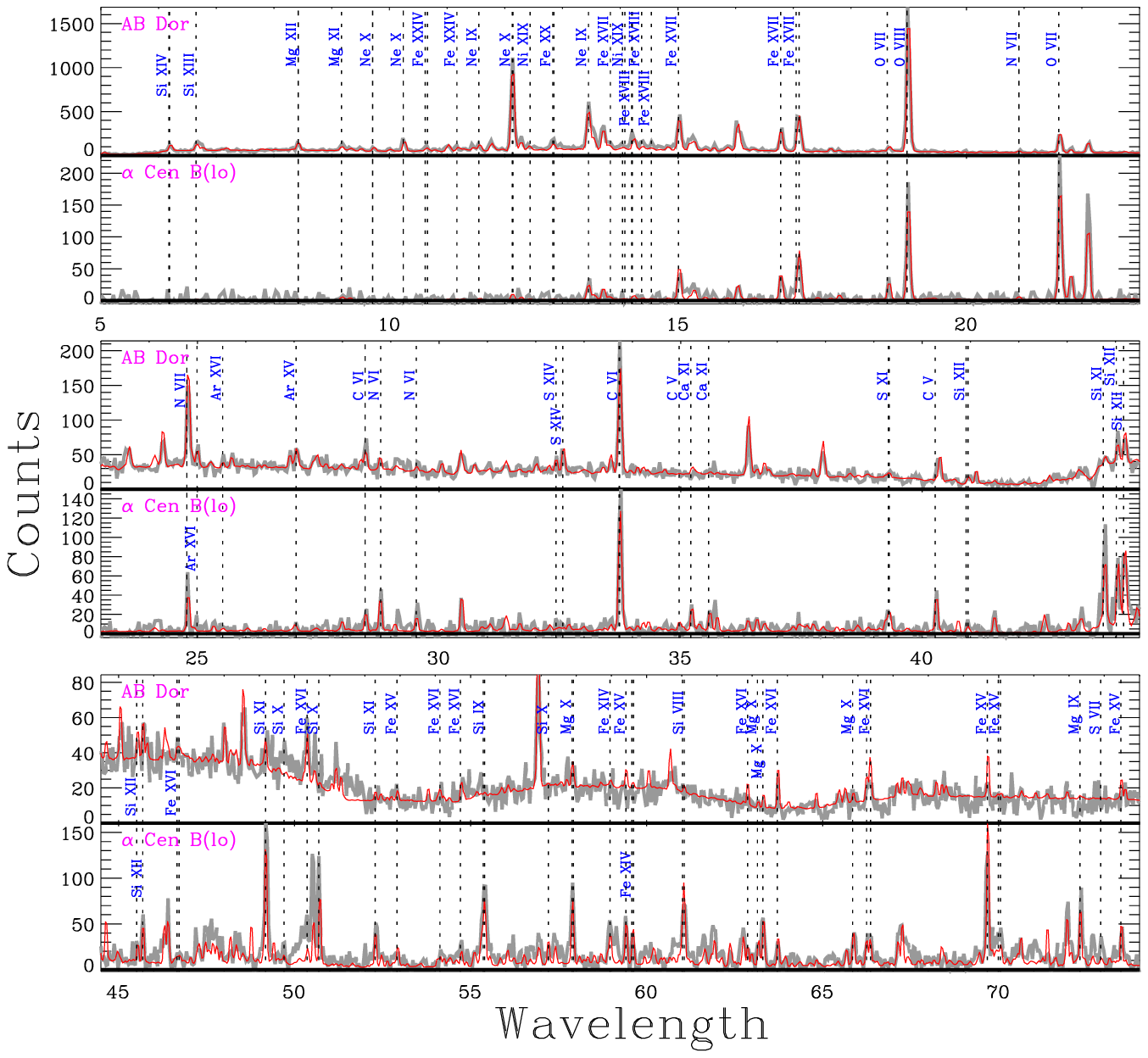}{4.7in}{0}{100}{100}{-300}{-367}
\caption{{\em Chandra}/LETGS spectra of AB~Dor~A and
  $\alpha$~Cen~B(lo), with the former representing a high-activity,
  high coronal temperature star; and the latter representing a low-activity,
  low coronal temperature star.  Lines used in the EM analysis are
  identified in the figure.  The red lines are synthetic spectra
  computed from the EM distributions derived from the spectra.}
\end{figure}
\setcounter{figure}{0}
\begin{figure}[t]
\plotfiddle{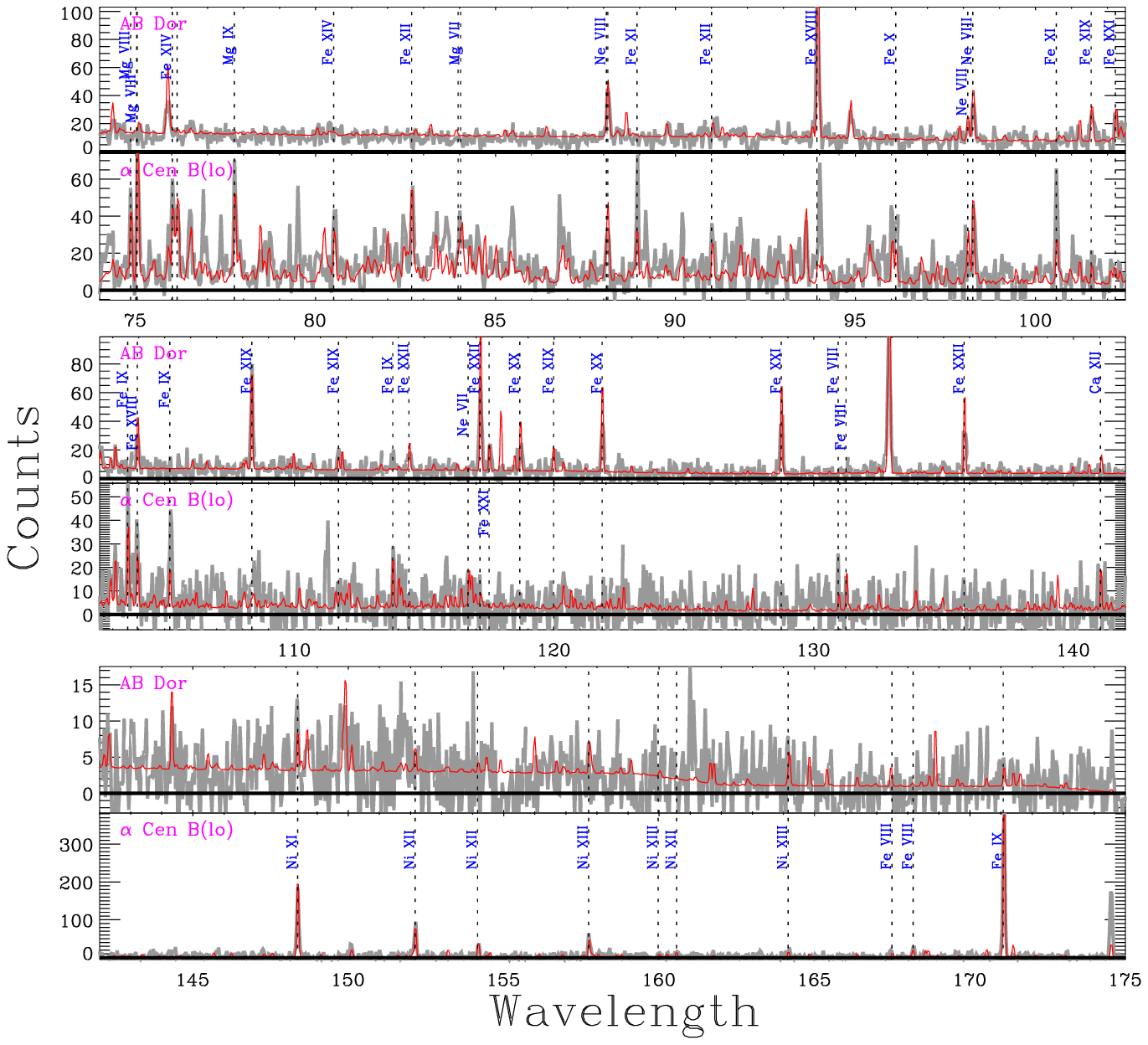}{4.7in}{0}{100}{100}{-300}{-367}
\caption{(continued)}
\end{figure}
     The two high-quality spectra used in the initial line identification
process are those of AB~Dor~A and $\alpha$~Cen~B(lo).  The former provides
a representative spectrum of a high-activity star with high coronal
temperatures, and the latter provides a representative spectrum of a
low-activity star with low coronal temperatures.  These spectra are
shown in Figure~1.

\begin{table}
\scriptsize
\begin{center}
\caption{Line Measurements$^a$}
\begin{tabular}{lcccccc} \hline \hline
Ion & $\lambda_{rest}$ & $\log T$ & \multicolumn{2}{c}{AB Dor A} & \multicolumn{2}{c}{$\alpha$ Cen B(lo)} \\
    &    (\AA)         &          & Counts & Flux ($10^{-5}$) & Counts & Flux ($10^{-5}$) \\
\hline
Si XIV & 6.180&7.32 & $160.1\pm 36.7$ & $2.95\pm 0.68$ & $<47.6$ & $<0.58$ \\
 & 6.186&  &   & $1.46\pm 0.33$ &   & $<0.57$ \\
Si XIII & 6.648&6.99 & $324.7\pm 42.6$ & $5.40\pm 0.71$ & $<46.2$ & $<0.54$ \\
... & ... & ... & ... & ... & ... & ... \\
Fe VIII & 168.173&5.84 & $<32.8$ & $<37.37$ & $59.0\pm 24.5$&$29.47\pm 12.24$ \\
Fe IX & 171.073&5.95 & $<33.4$ & $<31.30$ & $891.8\pm 43.7$&$366.38\pm 17.95$ \\
\hline
\end{tabular}
\end{center}
\tablecomments{$^a$Fluxes listed in units of
  $10^{-5}$ photons cm$^{-2}$ s$^{-1}$.  The complete version of
  this table is available online.}
\normalsize
\end{table}
     We have already published line lists and spectral analyses for a
number of stars in our sample \citep{bew06,bew10,bew13}.
But we here consider far more lines, many of
which are only identifiable now with better atomic data.  Our primary
line identification tool is version 7.1 of the CHIANTI atomic database
\citep{kpd97,el12,el13}.  However, published line
lists for the LETGS spectrum of Procyon are also valuable for
identifying lower temperature lines \citep{ajjr02,pb14}.
Table~3 provides our final list of
118 identified lines.  These lines are also noted in Figure~1.
A line formation temperature is estimated in the third column of the
table, based on a mean temperature of the line contribution function.
Also listed in Table~3 are counts
measured for each line and for each star.  Upper limits are measured
for nondetections.  These are $2\sigma$ limits computed from a
conservatively broad 0.19~\AA\ wavelength region around the line.
Note that the print version of Table~3 is an abbreviated table listing
only a few of the measurements for AB~Dor~A and $\alpha$~Cen~B(lo).
The full version of the table with all the line measurements for all
of the stars is available online.

     Given that we are only interested in lines that can be used
in our EM analysis, Table~3 does not list any emission features
that are blends of lines of different species.  In order to be
considered here, we have to believe that a line is at least
$\sim 80$\% from a single species, as ultimately verified using
synthetic spectra computed from the EM distributions that we
derive.  Examples of blends visible in
Figure~1 that are not considered are the O~VIII+Fe~XVIII blend
at 16.0~\AA\ and the Fe~XX+Fe~XXIII blend at 132.9~\AA.  We do
naturally consider blends of lines of a single species, as we can
use the line strengths in CHIANTI to divide the measured counts
among the individual lines in the blend.  Converting the line
counts listed in Table~3 to line fluxes, which are also listed
in the table (in units of photons~cm$^{-2}$~s$^{-1}$), involves
not only dividing the counts by exposure time and effective area,
but also dividing the counts into the various individual lines in
the case of blends.  This is why, for example, a single count
measurement is listed for the Si~XIV 6.2~\AA\ line, but two
flux measurements are listed for it, as we have divided the counts
between the $\lambda_{rest}=6.180$~\AA\ and $\lambda_{rest}=6.186$~\AA\
lines in the blend based on the CHIANTI emissivities.

     Perhaps the most extreme single-species line blend is the Fe~XX
line at 12.8~\AA, with dozens of Fe~XX lines that are near that
wavelength, which can be considered part of the blend.  In such cases,
our policy is to only list the two strongest lines in Table~3, and to
only consider those two lines in the EM analysis.  In the case of the
Fe~XX line, this means listing only the
$\lambda_{rest}=12.827$~\AA\ and $\lambda_{rest}=12.845$~\AA\ lines,
even though these two lines account for less than half the flux of the
blend according to the CHIANTI emissivities.  However, considering
more than two lines in a blend would give the blend too much weight in
the EM analysis, considering that a blended line is ultimately only a
single detected emission feature.  We experimented with
considering blends as only a single feature in the EM analysis, and
found no significant change in our results.  For the unresolved
density-sensitive He-like triplets Si~XIII $\lambda$6.7 and Mg~XI
$\lambda$9.2, we only list the strongest line in Table~3.

     In identifying lines in the various spectra, it is necessary to
be aware of higher order lines.  In LETGS spectra, the higher orders
are superposed on the first order spectrum, although even orders are
suppressed somewhat.  For example, in the AB~Dor~A spectrum the features
at 36.4~\AA\ and 56.9~\AA\ are third-order Ne~X $\lambda$12.1 and
O~VIII $\lambda$19.0, respectively.  The feature at
40.3~\AA\ is not C~V at $\lambda_{rest}=40.268$~\AA\ as Figure~1 might
seem to suggest, but is instead
third-order Ne~IX $\lambda$13.5, at least for AB~Dor~A.

     A few lines are worthy of brief discussion, starting with the
feature at 35.7~\AA.  In past analyses, using older versions of CHIANTI,
we identified this as S~XIII \citep{bew06,bew10,bew13}.
However, with CHIANTI version 7.1 this
identification no longer seems to work.  Nearby, there is a Ca~XI line
at 35.6~\AA\ that is now in our line list, but we are no longer sure
what the stronger 35.7~\AA\ emission is, so it is not identified in
Table~3 or Figure~1.

     Another problematic line is at 94~\AA.  For the active
stars that represent most of our sample, there is a line
centered at 93.9~\AA, which is clearly Fe~XVIII at
$\lambda_{rest}=93.932$~\AA.  However, for the inactive
$\alpha$~Cen~AB stars there is instead a peak at 94.0~\AA.  This is
clearly not Fe~XVIII, although for $\alpha$~Cen~B(hi) there is a
marginal detection of Fe~XVIII in the blue wing of the stronger
94.0~\AA\ line.  Given the numerous
Fe~X lines in the spectral region, the suspicion is that the 94.0~\AA\
line is Fe~X, but the version 7.1 CHIANTI line emissivities fail to
provide an unambiguous identification, given that synthetic spectra
fail badly to account for the line (see Figure~1).
\citet{pt12} discuss the difficulties in modeling this
feature in the LETGS spectrum of Procyon.
This is an important feature for solar physics, as the Atmospheric
Imaging Assembly (AIA) instrument on the {\em Solar Dynamics Observatory}
(SDO) mission includes a filter bandpass centered at 94~\AA, allowing
the Sun to be monitored at this wavelength.  Procedures for removing
the low temperature emission to determine a pure Fe~XVIII image have
been developed \citep{hpw12}.  These should work even
though current line emissivities cannot reproduce disk-integrated
spectra of this region from the Extreme
ultraviolet Variability Experiment (EVE) on SDO very well
\citep{sjs17}.

     Finally, one of the strongest lines in the $\alpha$~Cen~B(lo)
spectrum in Figure~1 is the Fe~X line at 174.5~\AA, which lies at
the very end of the LETGS spectral range.  However, the LETGS
effective area is falling rapidly at this wavelength, and our
experience suggests that the effective areas estimated at this
wavelength by the CIAO data reduction software are
unreliable.  Thus, we cannot include this line in our analysis.

\begin{figure}[t]
\plotfiddle{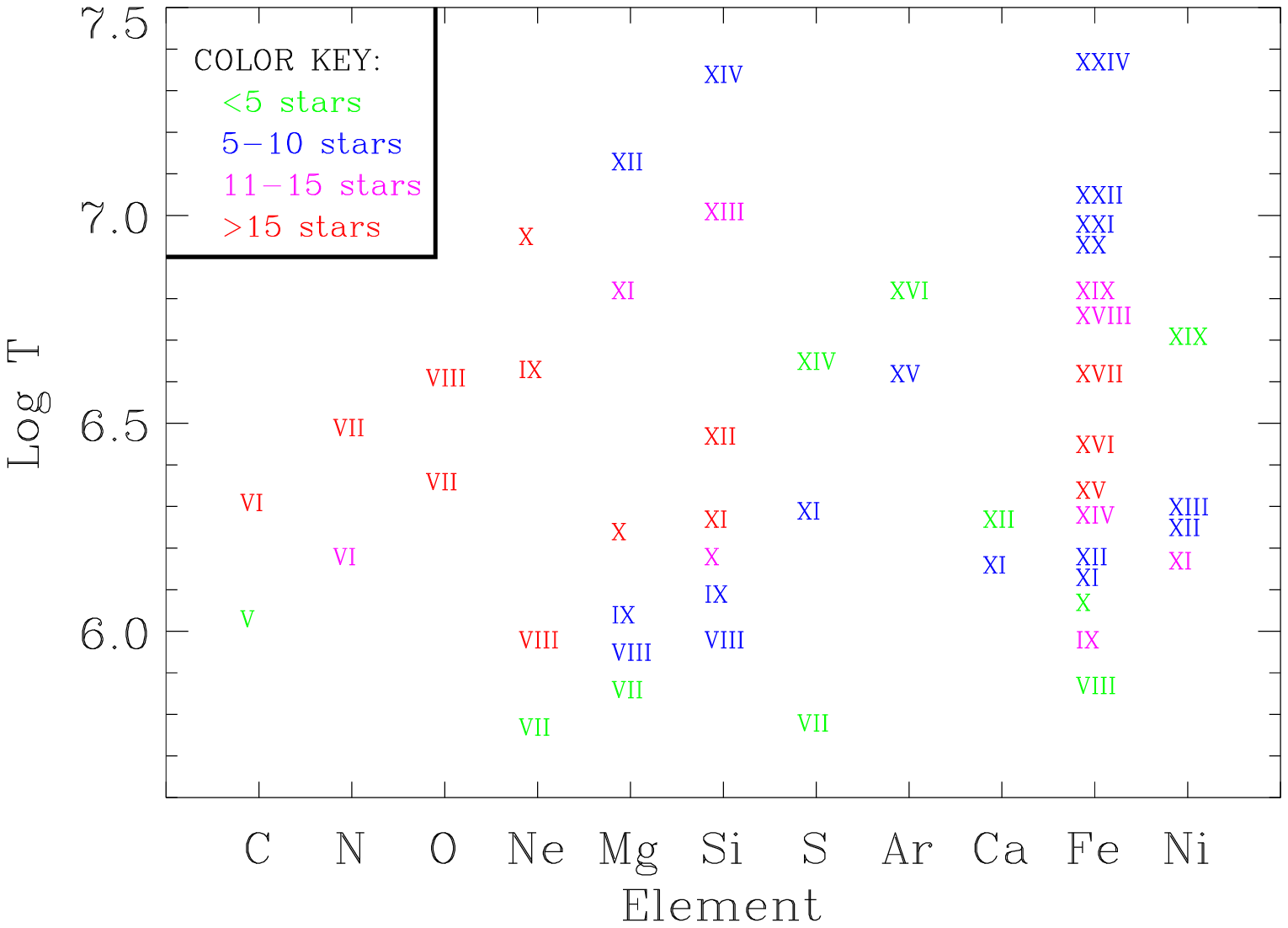}{3.2in}{0}{75}{75}{-250}{-285}
\caption{Illustration of the ionic species with detected emission lines
  in our sample of main sequence star {\em Chandra}/LETGS spectra.
  The species are plotted versus line formation temperature.
  A color scale is used to indicate the number of
  stars for which a given species has a detected line.}
\end{figure}
     Figure~2 provides a graphical illustration of the various ionic
species that are represented in our line list (e.g., Table~3), and how
many of them are detected in our sample of stars.  The detected lines
cover a range of line formation temperatures from $\log T=5.7-7.4$.
For Mg and Si, every species from Mg~VII--Mg~XII and Si~VIII--Si~XIV
is represented in the list.  For the particularly important Fe
sequence, every species from Fe~VIII---Fe XXIV is represented, except
for Fe~XIII and Fe~XXIII.  Technically, there are even detected lines
of Fe~XIII and Fe~XXIII at 76.5~\AA\ and 132.9~\AA, respectively (see
Figure~1), but these are blends with lines of other species and so do
not make our list.  Needless to say, not every species is detected for
every star.  The species in red in Figure~2 are detected in nearly
every spectrum, while those in green are only detected for a few.

\begin{figure}[t]
\plotfiddle{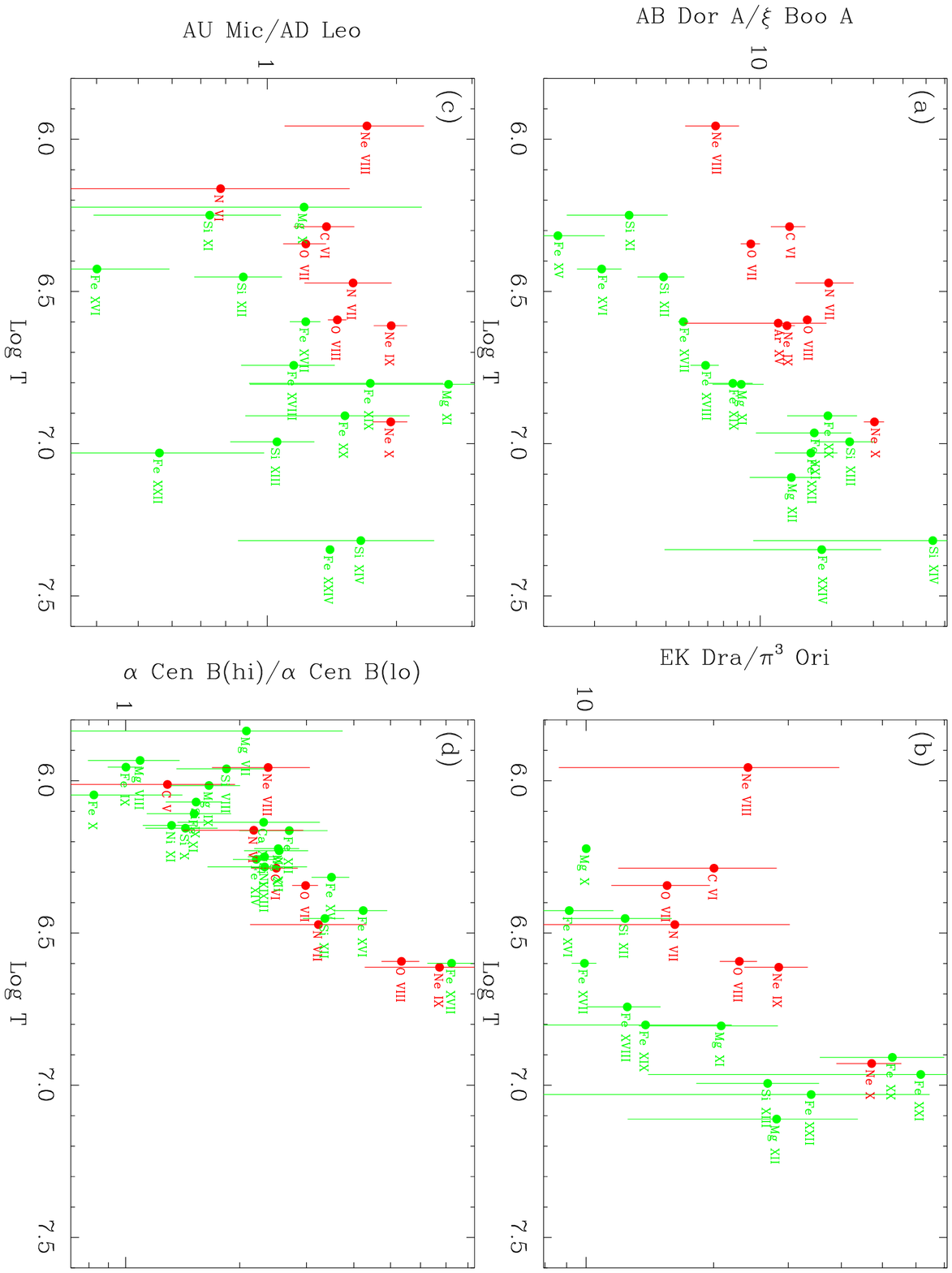}{5.1in}{90}{80}{80}{313}{-60}
\caption{The line fluxes of various stars are compared with each other
  by plotting line flux ratios versus the line formation temperature,
  with red (green) points indicating high-FIP (low-FIP) elements.}
\end{figure}
     A surprising amount can be learned from the line flux measurements
simply by comparing the fluxes of different stars with each other,
without the need for the complexity of the EM analysis that will be
described in the next section.  With 21 spectra in our sample,
there are lots of pairs of stars that can be compared.  Figure~3
shows examples of four such comparisons, plotting line flux ratios
versus line formation temperature, with different colors indicating
high-FIP (${\rm FIP}>10$ eV) and low-FIP (${\rm FIP}<10$ eV) elements.
We exclude sulfur lines in these figures, as S is on the border
between low-FIP and high-FIP.  The flux ratios include the obvious
corrections for differences in exposure time, distance, and radius;
but there is no attempt to correct for different reference photospheric
abundances.  In cases where there is more
than one line of a given species, we simply add fluxes of all lines
detected for both stars before computing the flux ratio.

     In all panels in Figure~3, the more active star with higher fluxes
is divided by the less active star with lower line fluxes.  Positive
slopes are seen in each panel, demonstrating that coronal temperatures
are higher for the more active stars.  Separation between the low-FIP
and high-FIP elements indicates a systematically different FIP effect.
Higher line ratios are clearly seen for high-FIP elements in
Figure~3(a-c).  This could be said to indicate a weaker FIP effect for
the more active stars, or alternatively a stronger inverse FIP effect.
The FIP bias seems to be relatively independent of temperature in the
sense that we find no evidence of high-FIP ratios being higher than
low-FIP ratios in one temperature range and lower in another
temperature range.  This is important because in the EM analysis
described in the next section, we have to assume uniform abundances
throughout the corona.  Figure~3(d) compares the ``(lo)'' and ``(hi)''
spectra for $\alpha$~Cen~B.  The increase in coronal temperature with
activity is very clear, but no evidence for any difference in FIP bias
is apparent.

\section{Emission Measure Analysis}

     With the line measurements made, the next step is an emission
measure analysis to infer coronal temperature distributions and
abundances.  This analysis requires assumptions of collisional
ionization equilibrium, Maxwellian velocity distributions, and uniform
abundances throughout the corona.  There are two definitions of
emission measure that will be used here, a volume emission measure,
$EM_V$ (in units of cm$^{-3}$), and a column emission
measure, $EM_h$ (in units of cm$^{-5}$).  For the former,
\begin{equation}
EM_V(T)\equiv n_e^2 \frac{dV}{d\log T},
\end{equation}
where $n_e$ is the coronal density and dV is a coronal volume element.
This is the most natural expression of emission measure for unresolved,
disk-integrated sources.  Note that we are expressing $EM_V$ as a
distribution in $\log T$, rather than $T$ as in some
EM definitions.  The observed flux for a given line can be
expressed as
\begin{equation}
f=\frac{1}{4\pi d^2} \int G(T,n_e)EM_V(T) d\log T,
\end{equation}
where $d$ is the stellar distance and $G(T,n_{e})$ is the line contribution
function, which includes both the line emissivity and the assumed
elemental abundance of the atomic species in question.

     For spatially resolved data (e.g., solar
observations), a column emission measure is more intuitive,
\begin{equation}
EM_h(T)\equiv n_e^2 \frac{dh}{d\log T}.
\end{equation}
This EM version is also more
useful when comparing emission measures of stars with different radii,
which we will be doing.  The distance element $dh$ can be related to
$dV$ by $dV=2\pi R^2 dh$, with $R$ the radius of the star.  We assume
$2\pi R^2$ instead of $4\pi R^2$ because the emission from the back side
of a star is hidden from us.  Thus,
\begin{equation}
EM_V=2\pi R^2 EM_h.
\end{equation}

     Knowing $EM_V$ allows line fluxes to be computed with relative
ease using equation (2).  However, the inverse problem of inferring
$EM_V$ from a set of emission line flux measurements is harder.
For this purpose, we use version 2.97 of the PINTofALE software developed
by \citet{vk00}, which includes a routine for computing emission measures
using a Markov Chain Monte Carlo approach
\citep{vk98}.  As in Section~3, version~7.1 of
the CHIANTI database is the source of our line emissivities.
The ionization equilibrium calculations of \citet{pm98} are
used, and we assume a typical coronal density of $\log n_e=10$.

\begin{table}
\scriptsize
\begin{center}
\caption{Coronal Abundance Measurements:  High FIP Elements}
\begin{tabular}{lcccccccc} \hline \hline
Star & $\log$ N$_H$ & [C/Fe] & [N/Fe] & [O/Fe] & [Ne/Fe] & [S/Fe] &
   [Ar/Fe] & $F_{bias}$ \\
\hline
$\eta$ Lep        & 17.87 & $0.78^{+0.27}_{-0.47}$ & ... & $1.09^{+0.16}_{-0.09}$ &
 $0.55^{+0.14}_{-0.19}$ & ... & ... & $-0.31\pm 0.09$ \\
$\pi^3$ Ori       & 17.93 & $0.79^{+0.10}_{-0.10}$ & $0.11^{+0.18}_{-0.35}$ &
 $0.92^{+0.05}_{-0.02}$ & $0.46^{+0.05}_{-0.06}$ & ... & $-0.86^{+0.46}_{-1.09}$ &
 $-0.41\pm 0.07$ \\
$\tau$ Boo A      & 17.82 & $0.83^{+0.29}_{-0.50}$ & ... & $0.91^{+0.16}_{-0.12}$ &
 $0.50^{+0.15}_{-0.34}$ & $0.11^{+0.38}_{-1.13}$ & ... & $-0.21\pm 0.09$ \\
$\pi^1$ UMa       & 18.12 & $0.54^{+0.37}_{-0.44}$ & $0.36^{+0.37}_{-0.70}$ &
 $0.84^{+0.17}_{-0.13}$ & $0.48^{+0.17}_{-0.22}$ & ... & ... & $-0.45\pm 0.18$ \\
EK Dra            & 18.08 & $0.98^{+0.21}_{-0.29}$ & $0.19^{+0.31}_{-0.63}$ &
 $1.12^{+0.11}_{-0.05}$ & $0.96^{+0.07}_{-0.07}$ & ... & ... & $0.00\pm 0.13$ \\
$\alpha$ Cen A(lo)& 17.61 & $0.72^{+0.08}_{-0.09}$ & $0.06^{+0.15}_{-0.36}$ &
 $0.35^{+0.15}_{-0.02}$ & $0.41^{+0.09}_{-0.09}$ & $0.07^{+0.09}_{-0.57}$ & ... &
 $-0.54\pm 0.30$ \\
$\alpha$ Cen A(hi)& 17.61 & $0.63^{+0.09}_{-0.10}$ & $0.08^{+0.18}_{-0.25}$ &
 $0.67^{+0.10}_{-0.11}$ & $0.33^{+0.06}_{-0.13}$ & $-0.12^{+0.12}_{-0.92}$ & ... &
 $-0.50\pm 0.15$ \\
$\xi$ Boo A       & 17.92 & $0.75^{+0.12}_{-0.06}$ & $0.20^{+0.15}_{-0.14}$ &
 $1.11^{+0.08}_{-0.03}$ & $0.69^{+0.07}_{-0.01}$ & $-0.45^{+0.10}_{-0.54}$ &
 $-0.46^{+0.18}_{-1.21}$ & $-0.30\pm 0.04$ \\
70 Oph A          & 18.06 & $0.64^{+0.16}_{-0.19}$ & $0.18^{+0.21}_{-0.30}$ &
 $0.92^{+0.11}_{-0.04}$ & $0.56^{+0.11}_{-0.16}$ & ... & ... & $-0.29\pm 0.06$ \\
AB Dor A          & 18.29 & $1.50^{+0.09}_{-0.04}$ & $1.06^{+0.08}_{-0.08}$ &
 $1.75^{+0.08}_{-0.03}$ & $1.35^{+0.05}_{-0.03}$ & $0.14^{+0.06}_{-0.33}$ &
 $-0.18^{+0.07}_{-0.37}$ & $0.60\pm 0.08$ \\
$\alpha$ Cen B(lo)& 17.61 & $0.70^{+0.09}_{-0.04}$ & $0.14^{+0.10}_{-0.14}$ &
 $0.87^{+0.04}_{-0.05}$ & $0.33^{+0.11}_{-0.13}$ & $-0.52^{+0.13}_{-0.27}$ & ... &
 $-0.38\pm 0.15$ \\
$\alpha$ Cen B(hi)& 17.61 & $0.70^{+0.05}_{-0.05}$ & $0.08^{+0.09}_{-0.10}$ &
 $0.89^{+0.04}_{-0.03}$ & $0.44^{+0.08}_{-0.08}$ & $-0.76^{+0.15}_{-0.23}$ & ... &
 $-0.36\pm 0.09$ \\
36 Oph A          & 17.85 & $0.57^{+0.39}_{-0.47}$ & $0.44^{+0.24}_{-0.71}$ &
 $1.19^{+0.15}_{-0.10}$ & $0.88^{+0.11}_{-0.24}$ & ... & ... & $-0.14\pm 0.10$ \\
36 Oph B          & 17.85 & $0.44^{+0.37}_{-0.36}$ & $0.33^{+0.34}_{-0.64}$ &
 $1.10^{+0.19}_{-0.13}$ & $0.67^{+0.24}_{-0.77}$ & ... & ... & $-0.27\pm 0.10$ \\
$\epsilon$ Eri    & 17.88 & $0.90^{+0.05}_{-0.04}$ & $0.42^{+0.06}_{-0.06}$ &
 $1.20^{+0.02}_{-0.02}$ & $0.84^{+0.03}_{-0.02}$ & $-0.45^{+0.15}_{-0.20}$ &
 $-0.42^{+0.09}_{-0.99}$ & $0.06\pm 0.07$ \\
$\xi$ Boo B       & 17.92 & $0.78^{+0.27}_{-0.26}$ & ... & $1.19^{+0.16}_{-0.13}$ &
 $0.97^{+0.12}_{-0.13}$ & ... & ... & $-0.16\pm 0.15$ \\
61 Cyg A          & 18.13 & $0.95^{+0.28}_{-0.21}$ & $0.34^{+0.40}_{-0.73}$ &
 $1.12^{+0.19}_{-0.10}$ & $0.79^{+0.39}_{-0.59}$ & $-0.20^{+0.43}_{-0.70}$ & ... &
 $-0.01\pm 0.04$ \\
70 Oph B          & 18.06 & $0.98^{+0.41}_{-0.15}$ & $0.70^{+0.33}_{-0.24}$ &
 $1.35^{+0.18}_{-0.08}$ & $1.01^{+0.14}_{-0.08}$ & ... & ... & $0.15\pm 0.10$ \\
61 Cyg B          & 18.13 & $1.26^{+0.38}_{-0.48}$ & ... & $1.60^{+0.22}_{-0.24}$ &
 $1.06^{+0.26}_{-0.29}$ & ... & ... & $0.33\pm 0.08$ \\
AU Mic            & 18.36 & $1.63^{+0.09}_{-0.08}$ & $1.19^{+0.09}_{-0.12}$ &
 $1.77^{+0.06}_{-0.06}$ & $1.39^{+0.05}_{-0.04}$ & ... & $-0.20^{+0.09}_{-1.62}$ &
 $0.68\pm 0.13$ \\
AD Leo            & 18.47 & $1.45^{+0.08}_{-0.08}$ & $0.96^{+0.09}_{-0.08}$ &
 $1.64^{+0.06}_{-0.04}$ & $1.18^{+0.06}_{-0.03}$ & $0.12^{+0.16}_{-0.43}$ & ... &
 $0.49\pm 0.11$ \\
\hline
\end{tabular}
\end{center}
\normalsize
\end{table}
     The PINTofALE routines include corrections for absorption from
the interstellar medium (ISM).  For stars as nearby as ours, ISM
column densities are low and the corrections modest, but they are
still well worth making for LETGS data, especially at higher
wavelengths above 100~\AA\ where the effects of ISM absorption become
more important.  The logarithmic ISM hydrogen column densities assumed
for our sample of stars, $\log N_H$ (in units of cm$^{-2}$), are
listed in Table~4.  The primary source for these columns is a
compilation of ISM column density measurements from {\em Hubble Space
Telescope} (HST) spectra of the H~I Lyman-$\alpha$ line \citep{bew05b}.
For $\pi^1$~UMa, the HST measurement is instead from \citet{bew14},
while for $\tau$~Boo~A $\log N_H$ is estimated
from the Mg~II column density of \citet{cm14} and the Mg
depletion for this direction from \citet{sr08}.  The
EK~Dra value of $\log N_H=18.08$ comes from a new H~I Lyman-$\alpha$
measurement that we make from an archival HST spectrum \citep{tra15}.
For details of how this kind of analysis is done, see \citet{bew05b}.
For $\eta$~Lep, $\pi^3$~Ori, and AB~Dor~A, no relevant ISM absorption
measurements exist, so we simply assume the $\log N_H$ values of stars
close to these targets in the sky.  Specifically, for $\eta$~Lep,
$\pi^3$~Ori, and AB~Dor~A the ISM columns listed in Table~4 are
actually those measured towards HD~43162, $\chi^1$~Ori, and
$\zeta$~Dor, respectively \citep{bew05b}.

\begin{figure}[t]
\plotfiddle{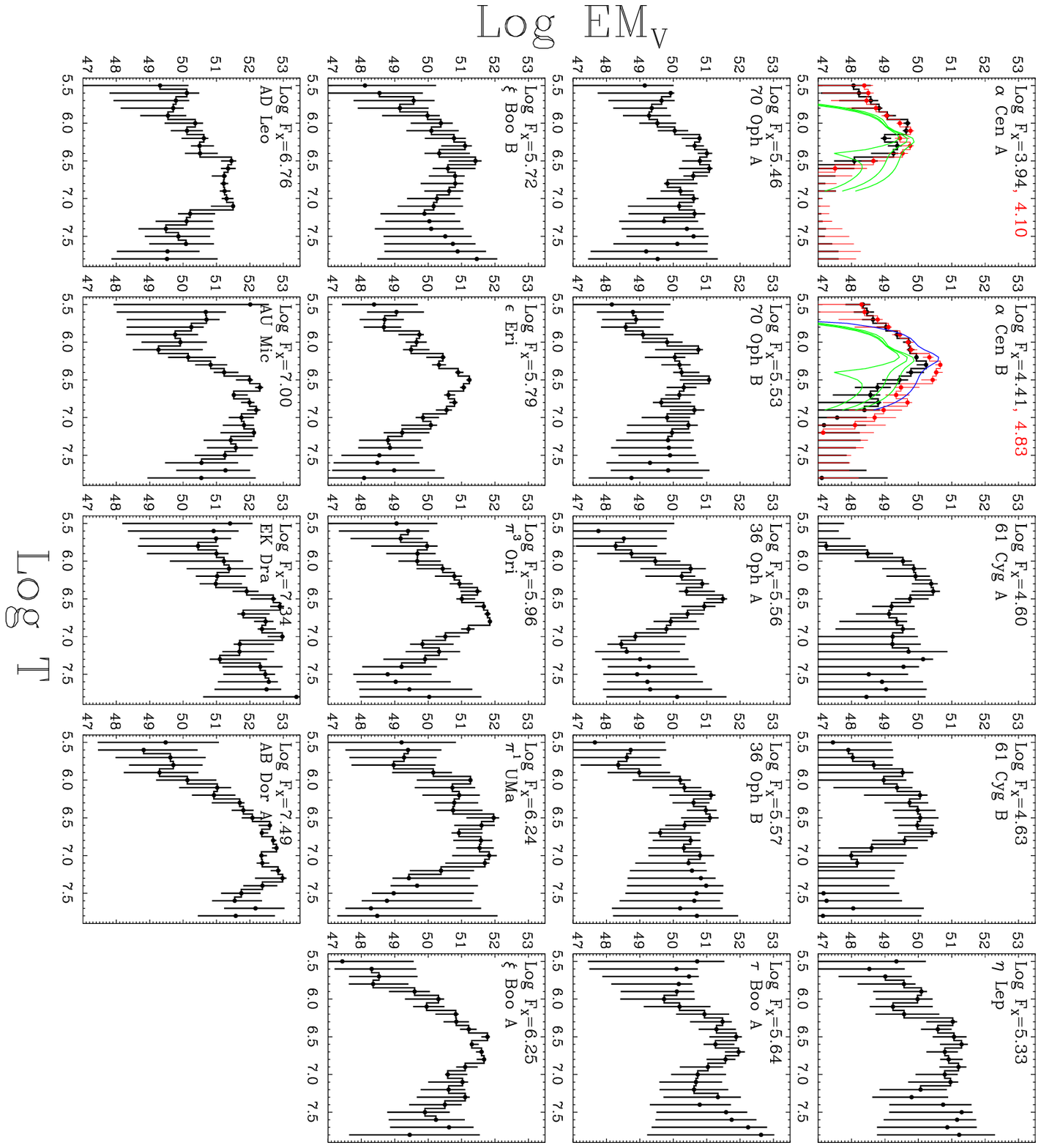}{5.6in}{90}{90}{90}{368}{-65}
\caption{Emission measure distributions computed from the main sequence
  star {\em Chandra}/LETGS spectra in our sample (units cm$^{-3}$).  The
  stars are in order of increasing $\log F_X$.  Dots indicate the best-fit
  $EM_V$ values.  The temperature range where a solid line connects
  these dots indicates the temperature range constrained by detected
  emission lines, with only upper limits contraining the EMs at
  other temperatures.  For $\alpha$~Cen~A and B, EMs for both
  the ``(lo)'' (black) and ``(hi)'' (red) spectra are shown.  Green
  lines indicate full-disk solar EM distributions measured at
  three different times \citep{sjs17}, with the blue
  line in the $\alpha$~Cen~B panel showing the intermediate solar
  distribution renormalized to assume the same coronal Fe abundance
  as measured for $\alpha$~Cen~B.}
\end{figure}
     Figure~4 shows the $EM_V$ distributions derived by the EM
analysis.  The EMs are computed from $\log T=5.5$ to $\log T=7.8$,
with a resolution of 0.1 dex.  Error bars are 90\% confidence
intervals suggested by the Monte Carlo analysis.  Technically, the
line-based EM analysis only determines the shape of the EM
distribution.  A line-to-continuum ratio analysis is required
to normalize it properly, which will be described below.  The temperature
range in which the best-fit EMs are connected by a solid line
provides an estimate of the temperature range actually
constrained by detected lines, which is estimated from the peak
temperatures of the line contribution functions, $G(T,n_e)$.
The lower temperature bound is defined by the lowest
peak temperature minus 0.3 dex, and the higher bound is the highest
peak temperature plus 0.3 dex.  Outside this range, the actual
constraints on EM are minimal.  The ``best-fit'' EM values and the
lower bounds indicated in the figure mean little outside this range,
but the upper bounds are well constrained because the EM analysis does
consider upper limits for all the undetected lines.

     The EMs in Figure~4 are shown in order of increasing activity level,
based on the logarithmic $F_X$ values indicated in the figure (in
erg~cm$^{-2}$~s$^{-1}$ units), which are computed from the radii and
X-ray luminosities in Table~1.  The general increase in both the
magnitude of $EM_V$ and the mean coronal temperature with $\log F_X$
is apparent.  For $\alpha$~Cen~A and B, both the
``(lo)'' and ``(hi)'' versions of the EM distribution are shown.

\section{Coronal Abundance Meaurements}

\begin{table}
\scriptsize
\begin{center}
\caption{Coronal Abundance Measurements:  Low FIP Elements}
\begin{tabular}{lccccc} \hline \hline
Star & [Mg/Fe] & [Si/Fe] & [Ca/Fe] & [Ni/Fe] & [Fe/Fe$_*$] \\
\hline
$\eta$ Lep        & $0.16^{+0.33}_{-0.75}$ & $-0.02^{+0.31}_{-0.74}$ & ... &
 $-0.56^{+0.04}_{-1.14}$ & (0.21)$^a$ \\
$\pi^3$ Ori       & $0.23^{+0.06}_{-0.10}$ & $0.24^{+0.07}_{-0.07}$ & ... &
 $-1.12^{+0.11}_{-0.27}$ & $-0.09^{+0.06}_{-0.07}$ \\
$\tau$ Boo A      & $0.22^{+0.27}_{-0.92}$ & $0.18^{+0.22}_{-0.54}$ & ... &
 ... & ($-0.62^{+0.08}_{-0.11}$)$^b$ \\
$\pi^1$ UMa       & $0.40^{+0.12}_{-1.09}$ & $0.14^{+0.24}_{-0.25}$ & ... &
 ... & ($0.24^{+0.10}_{-0.12}$)$^b$ \\
EK Dra            & $0.23^{+0.11}_{-0.28}$ & $0.26^{+0.08}_{-0.15}$ & ... &
 ... & $-0.30^{+0.06}_{-0.06}$ \\
$\alpha$ Cen A(lo)& $0.10^{+0.06}_{-0.07}$ & $0.08^{+0.08}_{-0.08}$ &
 $-0.74^{+0.11}_{-1.00}$ & $-1.44^{+0.09}_{-0.11}$ & (-0.03)$^a$ \\
$\alpha$ Cen A(hi)& $0.27^{+0.06}_{-0.09}$ & $0.15^{+0.09}_{-0.07}$ &
 $-0.77^{+0.15}_{-0.60}$ & $-1.30^{+0.11}_{-0.14}$ & (-0.03)$^a$ \\
$\xi$ Boo A       & $0.27^{+0.07}_{-0.05}$ & $0.10^{+0.09}_{-0.05}$ & ... &
 ... & $0.05^{+0.03}_{-0.03}$ \\
70 Oph A          & $0.30^{+0.17}_{-0.16}$ & $0.11^{+0.12}_{-0.09}$ & ... &
 ... & (-0.16)$^a$ \\
AB Dor A          & $0.18^{+0.13}_{-0.10}$ & $0.11^{+0.05}_{-0.05}$ & ... &
 $-1.06^{+0.03}_{-1.04}$ & $-0.64^{+0.01}_{-0.01}$ \\
$\alpha$ Cen B(lo)& $0.26^{+0.06}_{-0.05}$ & $0.03^{+0.10}_{-0.02}$ &
 $-0.91^{+0.18}_{-0.46}$ & $-1.15^{+0.11}_{-0.07}$ & $-0.61^{+0.11}_{-0.14}$ \\
$\alpha$ Cen B(hi)& $0.26^{+0.07}_{-0.05}$ & $0.11^{+0.04}_{-0.06}$ &
 $-1.02^{+0.04}_{-0.32}$ & $-1.16^{+0.07}_{-0.08}$ & $-0.61^{+0.05}_{-0.05}$ \\
36 Oph A          & $0.63^{+0.18}_{-1.29}$ & $0.19^{+0.21}_{-0.40}$ & ... &
 ... & $-0.25^{+0.15}_{-0.23}$ \\
36 Oph B          & $0.19^{+0.41}_{-0.77}$ & $0.01^{+0.35}_{-0.58}$ & ... &
 $-1.27^{+0.71}_{-0.81}$ & (-0.16)$^a$ \\
$\epsilon$ Eri    & $0.28^{+0.05}_{-0.06}$ & $0.12^{+0.05}_{-0.04}$ &
 $-0.87^{+0.23}_{-0.79}$ & $-1.19^{+0.19}_{-0.21}$ & $-0.15^{+0.03}_{-0.03}$ \\
$\xi$ Boo B       & $0.32^{+0.36}_{-1.08}$ & $0.19^{+0.23}_{-0.33}$ & ... &
 $-0.88^{+0.45}_{-0.96}$ & $-0.30^{+0.11}_{-0.15}$ \\
61 Cyg A          & $0.16^{+0.44}_{-0.78}$ & $0.15^{+0.36}_{-0.17}$ & ... &
 $-0.99^{+0.49}_{-1.11}$ & (-0.34)$^a$ \\
70 Oph B          & $0.55^{+0.25}_{-1.19}$ & $0.24^{+0.23}_{-0.51}$ & ... &
 ... & $-0.31^{+0.16}_{-0.25}$ \\
61 Cyg B          & $0.85^{+0.20}_{-1.28}$ & $0.53^{+0.37}_{-0.53}$ & ... &
 ... & $-0.61^{+0.12}_{-0.16}$ \\
AU Mic            & $0.36^{+0.10}_{-0.44}$ & $0.49^{+0.11}_{-0.10}$ & ... &
 ... & $-0.71^{+0.03}_{-0.03}$ \\
AD Leo            & $0.13^{+0.15}_{-0.23}$ & $0.53^{+0.07}_{-0.08}$ & ... &
 $-1.01^{+0.21}_{-0.73}$ & $-0.47^{+0.04}_{-0.04}$ \\
\hline
\end{tabular}
\end{center}
\tablecomments{$^a$Assumed value based on relation in Figure~7(c).
  $^b$Measurement from XMM/RGS data instead of {\em Chandra}/LETGS
  \citep{at05,am11}.}
\normalsize
\end{table}
     In the EM analysis, the abundances of elements with lines
of detected species are free parameters of the fits.  However,
the line-based analysis can only measure relative abundances, not
the absolute abundances (i.e., the abundances relative to the
dominant element, H).  Given the prevalence of Fe lines in our
spectra, Fe is the most obvious reference element to use for
quoting relative abundances.  Thus, the coronal
abundances relative to Fe measured by the EM analysis are listed
in Tables~4 and 5, with high-FIP elements listed in the former
and low-FIP elements listed in the latter.  We follow the
common convention where abundances surrounded by square brackets
are logarithmic, so the abundances in Tables 4 and 5 are listed in
logarithmic form.  Error bars are 90\% confidence intervals, as in
Figure~4.

     The scattering processes that dominate continuum emission at
X-ray wavelengths depend mostly on the abundances of H and He
\citep[e.g.,][]{jjd98}, so the line-to-continuum ratio provides a measure
of absolute abundances.  With all coronal abundances measured relative
to Fe from the emission line analysis,
determining absolute abundances reduces to measuring the
absolute abundance of Fe, i.e.\ [Fe/H].  We experimented with a number
of different ways to do the line-to-continuum analysis, using various
wavelength regions.  We even explored a sophisticated approach
of allowing the considered wavelength regions to be different for
different stars depending on where the highest continuum S/N is
after line subtraction.  However, this more complex approach did
not seem to lead to any clear advantage, so for the sake of
simplicity we ultimately focus exclusively on the
$25-40$~\AA\ region, which is relatively devoid of strong emission
lines, except for C~VI $\lambda$33.7, but is still at wavelengths short
enough where the continuum is stronger.

\begin{figure}[t]
\plotfiddle{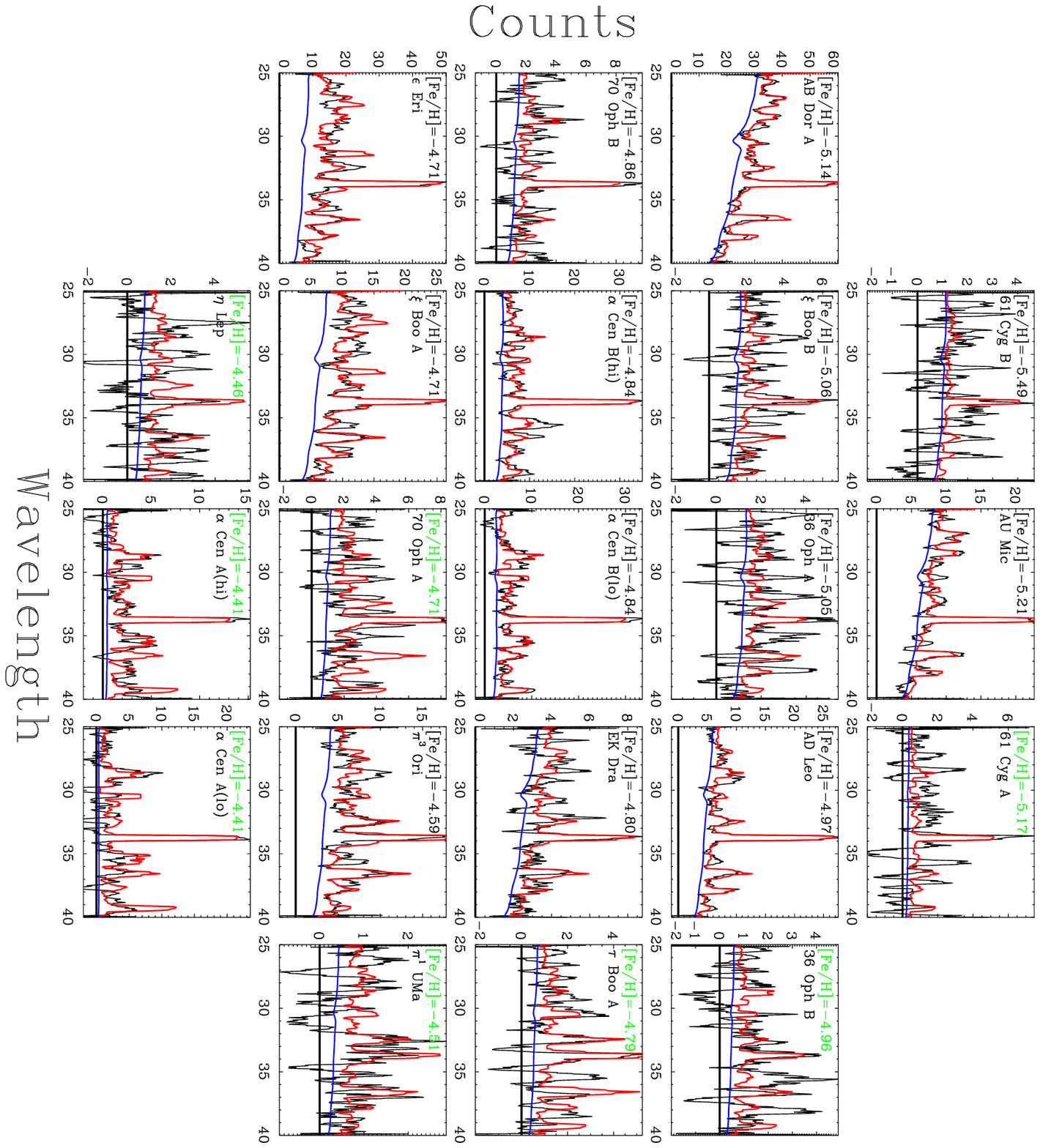}{5.6in}{90}{90}{90}{368}{-65}
\caption{Highly smoothed LETGS spectra of the $25-40$~\AA\
  region, which are used to measure the continuum level and determine
  absolute coronal abundances.  The zero flux level is indicated by
  a horizontal black line.  Blue lines indicate the inferred
  continuum level, based on the EM distributions in Figure~4.  The
  red lines are the final synthetic spectra after the inclusion of
  the line emission.  The absolute iron abundance suggested by the
  continuum level, [Fe/H], is indicated in each panel, with the
  panels arranged in order of increasing [Fe/H].  Values in green
  are assumed rather than measured, as these are cases where
  we do not have a statistically significant detection of the
  continuum.}
\end{figure}
     Figure~5 shows the LETGS $25-40$~\AA\ spectra.  Our procedure
for measuring the continuum level here involves first measuring the
total flux in this range (skipping a narrow wavelength region around
the C~VI line), and then subtracting from this flux the integrated
emission line flux inferred from a synthetic $25-40$~\AA\ spectrum
computed using the EM distribution in Figure~4.  This provides a
continuum flux estimate, $f_{cont}$.  Using the relevant PINTofALE
procedures, we determine the absolute Fe abundance, [Fe/H],
necessary to yield a continuum that reproduces $f_{cont}$.  Figure~5
shows both the continuum-only and total line-plus-continuum synthetic
spectra.

    The inferred [Fe/H] values are indicated in Figure~5.
The absolute Fe abundances are also listed in Table~5, but in the
table we list them as [Fe/Fe$_*$], in order to indicate the coronal
Fe abundance relative to the stellar photospheric Fe abundance
from Table~1.  For each $f_{cont}$ measurement, we also measure
a Poissonian uncertainty based on the noise of the spectrum, and
for eight of our spectra we find $f_{cont}$ is less than $2\sigma$
above the noise.  For these spectra, we conclude that we do not have a
statistically significant detection of the continuum, meaning that
we cannot make a meangful measurement of [Fe/H].  For these eight
cases, we have to assume [Fe/H] by other means.  The assumed
values are flagged in Table~5 and colored green in Figure~5.

     For $\tau$~Boo~A and $\pi^1$~UMa the solution is to simply use
published [Fe/H] measurements from XMM \citep{at05,am11}.
Observations from XMM are well suited for
continuum measurement, with the high sensitivity of RGS and the
possibility of considering spectra from both RGS and
from the EPIC instrument.  Pulse height spectra from EPIC do not
provide much spectral resolution, but this is far less important
for continuum measurement than for line measurement.  In
contrast, LETGS is not ideal for continuum measurements, as the high
background of the HRC-S detector can make it difficult to detect weak
continua.  For emission lines, this problem is mitigated somewhat by
the excellent spectral resolution of LETGS, which means the emission
is isolated to a limited number of pixels with therefore a limited
number of background counts.  But high spectral resolution is less
helpful for detecting weak continuum emission.

     This still leaves us with six spectra with no [Fe/H] measurement
[61~Cyg~A, 36~Oph~B, 70~Oph~a, $\eta$~Lep, $\alpha$~Cen~A(hi), and
$\alpha$~Cen~A(lo)].  Estimates of some sort are required, as
the [Fe/H] values are not only of interest in their own right, but are
also necessary to normalize the EM distributions.  Increasing [Fe/H]
would correspond to lowering the $EM_V$ curves in Figure~4.  We will
use our existing [Fe/H] measurements to estimate
[Fe/H] for the stars without a measurement, but before
describing that in detail we first discuss the $F_{bias}$ values
listed in the final column of Table~4.

\begin{figure}[t]
\plotfiddle{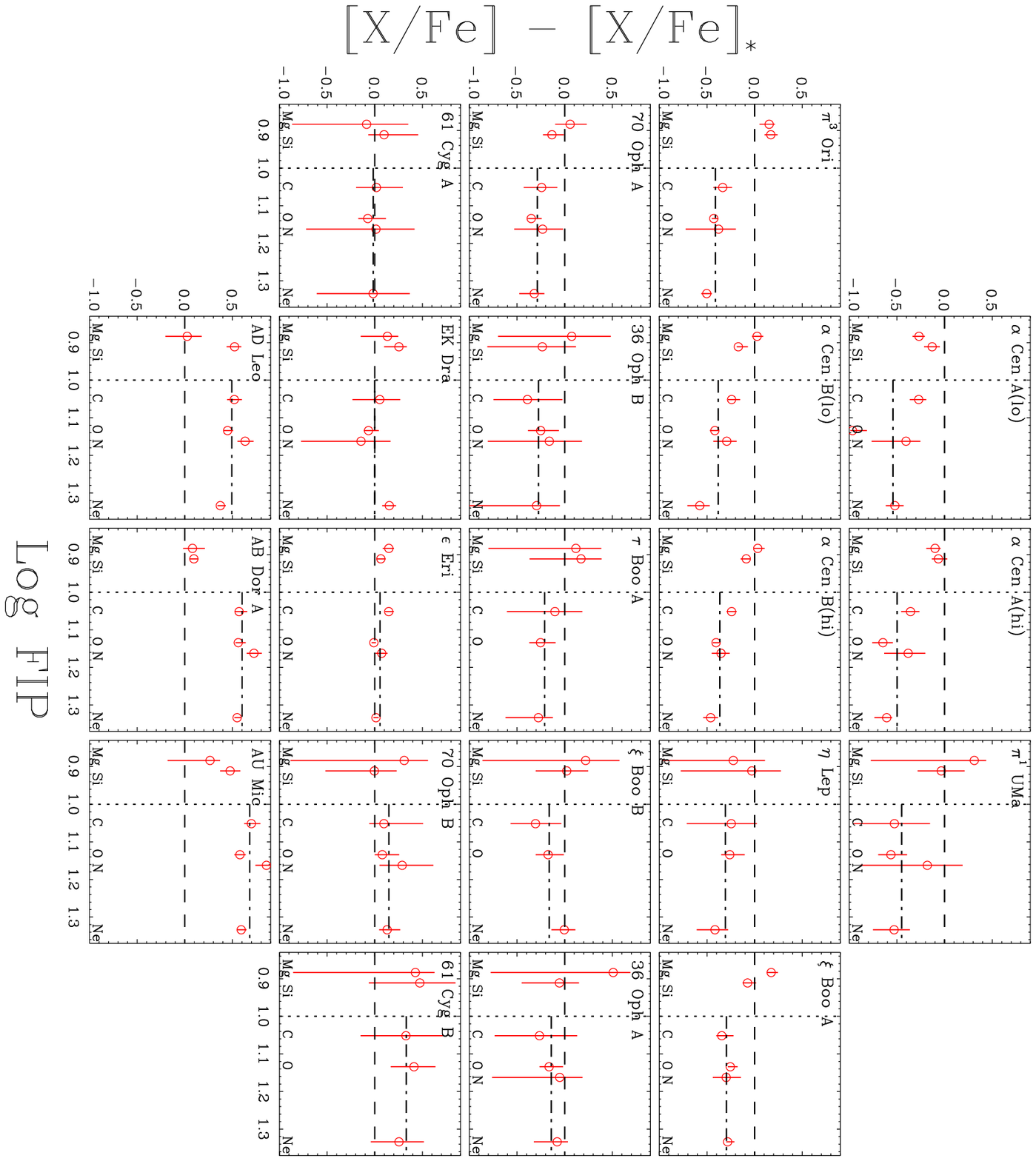}{5.6in}{90}{90}{90}{368}{-65}
\caption{Coronal abundances of six elements relative to Fe, [X/Fe],
   versus FIP (in eV), relative to the stellar photospheric ratio,
  [X/Fe]$_*$.  A vertical dotted line separates the low-FIP elements
  (Mg and Si) from the high-FIP elements (C, N, O, and Ne).  The dot-dashed
  line is the average value of the four high-FIP elements, which is the
  $F_{bias}$ parameter listed in Table~4.  The stars are shown in order of
  increasing $F_{bias}$.}
\end{figure}
     The ``FIP-bias'' parameter, $F_{bias}$, represents an attempt to 
define a simple metric that can be used to quantify
a star's coronal abundance characteristics \citep{bew10}.
It is the average logarithmic
abundance of four high-FIP elements (C, N, O, and Ne) relative to
Fe, which is the best-constrained low-FIP element.  This abundance ratio
is computed relative to stellar photospheric abundances.  Figure~6
shows explicitly how $F_{bias}$ is derived for our sample of
LETGS-observed stars.  In the figure, the abundance ratios of the
six best-measured coronal abundances relative to Fe ([X/Fe] from
Tables~4 and 5) are plotted versus FIP (in eV), with a correction for
the stellar photospheric abundance ratios, [X/Fe]$_*$, from Table~1.
Thus, a value of 0.0 corresponds to a coronal abundance ratio identical
to that of the stellar photosphere.  

     The abundance ratios in Figure~6 use the stellar photospheric
abundance figures in Table~1, which must be converted to absolute
abundances by assuming solar reference abundances.  For that purpose,
we use \citet{ma09}.  No photospheric abundances are listed in
Table~1 for N or Ne, as no stellar measurements exist for these
elements.  For N, we simply use O as a proxy, assuming
[N$_*$/N$_{\odot}$]=[O$_*$/O$_{\odot}$], with the N abundance ratios
in Figure~6 reflecting this assumption.  The situation for Ne is more
complicated, as no direct photospheric measurements are possible even
for the Sun, due to a lack of photospheric Ne lines.  Solar Ne
abundances are instead inferred from coronal and transition region
lines.  The Ne abundance is often quoted relative to O, another
high-FIP element.  The \citet{ma09} tabulation suggests
Ne/O=0.17, but much higher Ne/O values are generally found for stellar
coronae.  In particular, \citet{jjd05} find Ne/O=0.41.  The
most recent solar measurement from \citet{pry18} finds Ne/O=0.24, which
is a higher value than reported before but still well below the
stellar value.  It remains debatable which
measurement is preferable to represent the reference photospheric Ne
abundance, but we follow past practice \citep[e.g.,][]{bew13}
and use the \citet{jjd05} Ne/O ratio.

     In Figure~6, $F_{bias}$ is simply the average value of the
four high-FIP elements, minus N in a few cases where no N lines
are detected.  This average is
shown explicitly in the figure, with the stars displayed in order of
increasing $F_{bias}$.  Values below 0.0 represent a solar-like FIP
effect, while values above 0.0 represent an inverse FIP effect.  We
compute a standard deviation of the four individual ratios, and use
that as an estimate of the uncertainty in $F_{bias}$.  The $F_{bias}$
values and their uncertainties are listed in Table~4.

     At this point, we can compare our EM distributions and $F_{bias}$
measurements with ones made in the past,
both from previous analyses of the same LETGS data studied here and from
analyses of other data (e.g., XMM).  We find two surprising differences.
The first is that our new $F_{bias}$ measurements are systematically
higher than before.  The mean and standard deviation of the difference
is $\Delta F_{bias}=+0.084\pm 0.052$.  (See Table~2 of \citet{bew12}
for a list of previous $F_{bias}$ measurements.)  The second
difference concerns the shape of the EM distributions.  A feature that
we emphasized in the past was a surprisingly sharp peak in the EM
distribution at $\log T=6.6$, which seemed ubiquitous and particularly
convincing for high S/N data \citep{bew10,bew13}.
However, in Figure~4 such sharp and narrow
peaks at $\log T=6.6$ do not exist.  There is still generally an
EM maximum near this temperature, but the exact $\log T=6.6$
temperature is no longer emphasized like it was before.

     We trace both changes to large adjustments in the
emissivities of Fe~XVII lines in version~7.1 of CHIANTI, compared to
older versions \citep{el12}.  There are four Fe~XVII lines
in the $15-17$~\AA\ range in our line list, which are usually the
strongest Fe lines in the entire LETGS spectral range.  These lines
have rest wavelengths of $\lambda_{rest}=15.013$~\AA, 16.776~\AA,
17.051~\AA, and 17.096~\AA; the last two being in an unresolved blend.
The emissivity of the $\lambda_{rest}=15.013$~\AA\ line changed little
from version~6 of CHIANTI, but the other three increased by an average
of 60\%.  The strength of these Fe~XVII lines makes it particularly
important for an EM analysis to reproduce these line fluxes.  With the
line emissivities being apparently underestimated in the past, the EM
reconstruction routine (PINTofALE in our case) would correct for this
in two ways: 1. Increase the Fe abundance (thereby decreasing
$F_{bias}$), and 2.  Create a sharp EM spike at $\log T=6.6$, which is
the peak of the Fe~XVII line contribution function.  Further
discussion of difficulties with Fe~XVII emissivities can be found
elsewhere \citep{pb04,gyl10,sb12}.
For our purposes, this serves as a reminder that
uncertainties in line emissivities remain a crucial factor limiting
the accuracy of EM analyses.

     In Figure~7(a), $F_{bias}$ is plotted versus spectral type,
reproducing the FBST relation discussed in Section~1.  In plots such
as this, it is worthwhile to supplement our {\em Chandra}/LETGS
measurements with other published results.  Table~6 lists 10 stars
with published X-ray spectral analyses, from which we can extract both
$F_{bias}$ and [Fe/Fe$_*$] values.  These include the Sun, for
which \citet{jts12} measures [Fe/Fe$_*$]=0.33 and
$F_{bias}=-0.48$.  Table~6 also includes an
LETGS $F_{bias}$ measurement for GJ~338~AB.  The GJ~338~AB spectrum
was not of sufficient quality for a full EM analysis, but
an $F_{bias}$ value could be estimated from the Fe~XVII/O~VIII line ratio
\citep{bew12}.  The other measurements are either from XMM/RGS or
{\em Chandra}/HETGS data.  We expect all the  published analyses
(excepting the Sun) will be affected by the aforementioned issue with
the Fe~XVII line emissivities.  Thus, based on our experience with the
LETGS sample, we have adjusted the $F_{bias}$ values
in Table~6 by $+0.084$.
\begin{figure}[t]
\plotfiddle{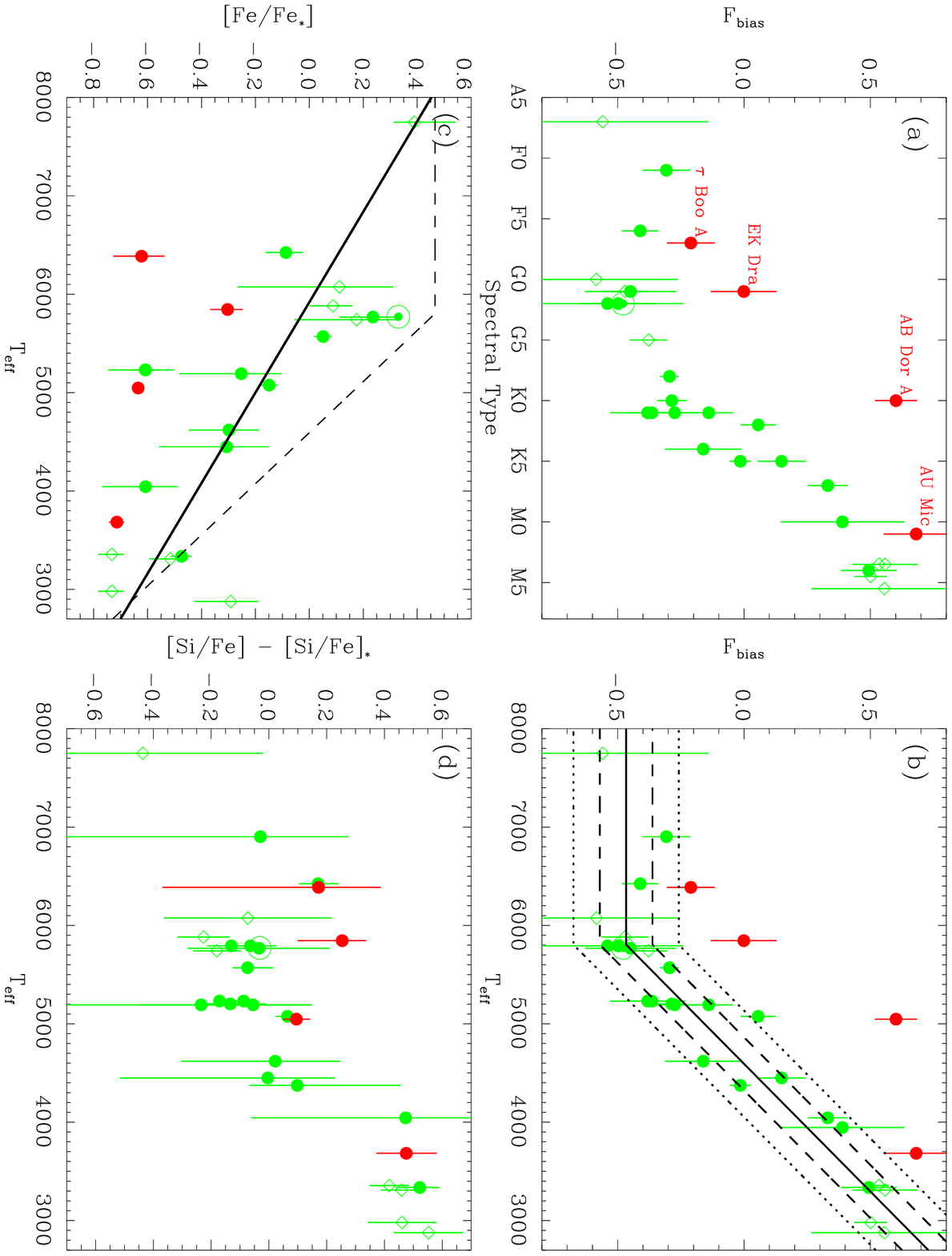}{5.1in}{90}{80}{80}{315}{-60}
\caption{(a) The FIP bias, $F_{bias}$, versus spectral type
  for the LETGS sample of stars in Table~1 (filled circles), and the
  literature values from Table~6 (diamonds).  We refer to the apparent
  correlation as the FIP-bias/spectral-type (FBST) relation.  Red
  symbols indicate four stars inconsistent with the FBST relation due
  to high activity or exoplanet effects.  (b) $F_{bias}$ versus
  $T_{eff}$, with a linear fit made to the points consistent with the
  FBST relation, assuming a flattening at higher temperatures.  Dashed
  and dotted lines indicate the $1\sigma$ and $2\sigma$ scatter about
  the best fit.  (c) Coronal Fe abundances relative to the photosphere
  are plotted versus $T_{eff}$.  A linear fit to the data is shown,
  excluding the four inconsistent stars noted above.  The dashed line
  is the expected relation if only the low-FIP elements are
  fractionated relative to H, based on the relation from (b).
  (d) The Si abundance relative to Fe (normalized to the photospheric
  ratio) is plotted versus $T_{eff}$. Note the high values seen at
  low $T_{eff}$.}
\end{figure}
\begin{table}[t]
\scriptsize
\begin{center}
\caption{Other Main Sequence Stars with Coronal Abundance Measurements}
\begin{tabular}{lccccccccccccc} \hline \hline
Star & Spectral & Dist. & Radius    & $T_{eff}$ & $\log L_X$  & [Fe/Fe$_*$] &
  $F_{bias}$$^{a}$ & Source & Refs. \\
     &   Type   &   (pc)   &(R$_{\odot}$)&  (K)    &(erg s$^{-1}$)& & & & \\
\hline
Altair       & A7 V & 5.13 & 1.93 & 7750 & 27.45 & $0.38^{+0.15}_{-0.08}$ &
 $-0.56\pm 0.42$ & XMM/RGS & 1 \\
$\beta$ Com  & G0 V & 9.13 & 1.11 & 6075 & 28.21 & $0.11^{+0.20}_{-0.38}$ &
 $-0.58\pm 0.32$ & XMM/RGS & 2 \\
$\chi^1$ Ori & G1 V & 8.66 & 0.98 & 5882 & 28.99 & $0.09^{+0.07}_{-0.09}$ &
 $-0.47\pm 0.09$ & XMM/RGS & 2 \\
Sun          & G2 V & ...  & 1.00 & 5771 & 27.35 & 0.33 &$-0.48$& ... & 3 \\
$\kappa$ Cet & G5 V & 9.14 & 0.92 & 5742 & 28.79 & $0.18^{+0.15}_{-0.23}$ &
 $-0.38\pm 0.07$ & XMM/RGS & 2 \\
GJ 338 AB &M0 V+M0 V& 5.81 &0.58+0.57&3946&27.92 & ... &
 $0.39\pm 0.25$  & LETGS   & 4 \\
EQ Peg A    &M3.5 Ve& 6.18 & 0.35 & 3356 & 28.71 & $-0.73^{+0.05}_{-0.05}$ &
 $0.53\pm 0.04$  & HETGS   & 5 \\
EV Lac      &M3.5 Ve& 5.12 & 0.30 & 3310 & 28.99 & $-0.52^{+0.07}_{-0.08}$ &
 $0.56\pm 0.13$  & XMM/RGS, HETGS   & 5,6 \\
EQ Peg B    &M4.5 Ve& 6.18 & 0.25 & 2981 & 27.89 & $-0.73^{+0.05}_{-0.05}$ &
 $0.50\pm 0.07$  & HETGS   & 5 \\
Prox Cen    &M5.5 Ve& 1.30 & 0.14 & 2877 & 27.22 & $-0.29^{+0.10}_{-0.13}$ &
 $0.55\pm 0.29$  & XMM/RGS, HETGS   & 5,7 \\
\hline
\end{tabular}
\end{center}
\tablerefs{(1) \citet{jr09}; (2) \citet{at05};
  (3) \citet{jts12}; (4) \citet{bew12}; (5) \citet{cl08};
  (6) \citet{jr05}; (7) \citet{mg04b}.}
\tablecomments{$^a$$F_{bias}$ values adjusted upward by $+0.084$
  (excluding the Sun) to correct for atomic data changes (see text).}
\normalsize
\end{table}

     As reported in past studies, we find that the relation
between $F_{bias}$ and spectral type is surprisingly tight for main
sequence stars (see Section~1).  However, this is the case only if
one ignores particularly active stars with $\log L_X>29$
(or $\log F_X>7$).  The three stars in our sample that fall in this
category (EK~Dra, AB~Dor~A, and AU~Mic) are identified in red in
Figure~7.  In all three cases, the points lie well above the FBST relation.
Evolved stars such as Procyon (F5~IV-V) and Capella (G8~III+G1~III)
also seem to universally lie above the FBST relation \citep{bew13}.
The $\tau$~Boo~A data point is also shown in red in Figure~7,
as it seems slightly high.  Assessing whether $\tau$~Boo's coronal
abundances are affected by the presence of its close-in,
massive exoplanet is one goal of our analysis (see Section~8).

     Aside from the systematic changes in $F_{bias}$ noted above,
there are two other notable changes to the FBST relation in Figure~7(a)
compared to past work.  One is the inclusion of error
bar estimates for $F_{bias}$.  The second and more important change is
the extension of the relation to A and early F spectral types.  For
the first time, we can definitively show that $F_{bias}$ values do not
continue to decrease beyond early G spectral types.  Solar-like
$F_{bias}$ values are seen from late A to early G.

     The assessment of the FBST relation for earlier type stars
is possible thanks to the consideration of the very recent
LETGS observation of $\eta$~Lep (F1~V) and the inclusion of the XMM
measurement for Altair (A7~V), from \citet{jr09}.  The
primary reason Altair had not been considered before was that Altair's
spectral type is sometimes listed as A7~IV-V.  Our experience with
Procyon (F5~IV-V) suggests that luminosity class IV-V stars can be
inconsistent with the FBST relation \citep{bew13}.  However, we
have now concluded that we can consider Altair a main sequence star
for our purposes, primarily because the $F_{bias}$
measurements of Altair and $\eta$~Lep provide a consistent picture of
an FBST relation that is flat at early spectral types (late A through
early G).  Altair's surface gravity is significantly higher than Procyon's,
and closer to that expected for the main sequence
\citep{mlm90}.  Altair's mean radius of
1.93~R$_{\odot}$ might be considered consistent with an A7~V spectral type,
considering the rapid rotation and resulting asphericity of this star
\citep{no04}.  \citet{dmp06} conclude that Altair
is close to the zero-age main sequence despite the IV-V classification.

     Figure~7(b) shows another version of the FBST relation,
with photospheric temperature
replacing spectral type on the x-axis.  This substitution makes it
easier for us to actually quantify the FBST relation.  We perform
a linear fit to the $F_{bias}$ measurements (ignoring the red data points),
with a flattening at a temperature that is another free parameter
of the fit.  The resulting relation is
\begin{equation}
F_{bias} = \left\{ \begin{array} {c @{\quad\mbox{for}\quad} l}
  1.77-3.85\times 10^{-4}~T_{eff} & T_{eff}<5804 \\
  -0.47 & T_{eff}>5804.
  \end{array} \right.
\end{equation}
The $1\sigma$ scatter of the $F_{bias}$ values about the best fit
is $\pm 0.104$.  Figure~7(b) shows the $1\sigma$ and $2\sigma$
deviations from the best fit.  The four red points that we have
considered inconsistent with the FBST relation (albeit only
tentatively for $\tau$~Boo~A) are indeed above the $2\sigma$ line,
although a single green point ($\epsilon$~Eri) is slightly above as
well.

     Switching from the relative abundances represented by $F_{bias}$
to the absolute abundances represented by [Fe/Fe$_*$], Figure~7(c)
plots [Fe/Fe$_*$] versus photospheric temperature.  If the four red
points are ignored, there seems to be a clear correlation, albeit
with more scatter than the $F_{bias}$ relations in Figures~7(a-b).
A linear fit yields
\begin{equation}
{\rm [Fe/Fe_*]}=-1.29+2.18\times 10^{-4}~T_{eff}.
\end{equation}
Unlike for the relative abundances, there is no evidence for
a flattening of the relation at early spectral types.
This conclusion, however, relies almost entirely on the
Altair measurement, and suffers from our inability to detect the
continuum and measure [Fe/Fe$_*$] for $\eta$~Lep.
We can now return to the issue of what to assume for [Fe/Fe$_*$] in
the six cases where we do not have an X-ray continuum detection 
[61~Cyg~A, 36~Oph~B, 70~Oph~a, $\eta$~Lep, $\alpha$~Cen~A(hi), and
$\alpha$~Cen~A(lo)].  For these six cases, we
simply use equation~(6) to estimate [Fe/Fe$_*$], and these are
the values listed in Table~5 for those stars.

     The four red points all lie below the relation in Figure~7(c).
While $\tau$~Boo~A is only marginally inconsistent with the FBST
relation in Figures~7(a-b), it is very inconsistent with the
relation in Figure~7(c), providing further support for the
star's coronal abundances being considered anomalous.  We
will discuss $\tau$~Boo further in Section~8.

     Aside from the $F_{bias}$ parameter, the individual element
abundance measurements in Tables~4 and 5 can be perused to search
for interesting behavior specific to particular elements.  One
example particularly worthy of note involves the
coronal Si abundances of M dwarfs.  In Figure~7(d), Si abundances
relative to Fe are plotted versus $T_{eff}$, with corrections for the
reference photospheric abundances.  For most stars, there is
no dramatic difference between the coronal and photospheric Si/Fe
ratio.  The exceptions are the $T_{eff}<4000$~K stars, i.e.\ the M
dwarfs, which collectively have high coronal Si/Fe.  The
various panels of Figure~7 imply that for M dwarfs, low-FIP elements
are coronally depleted by a factor of $3-4$, while the high-FIP
element abundances are roughly photospheric.  The exception is the
low-FIP Si, which seems to behave more like a high-FIP element, with
roughly photospheric coronal abundances.

     One of the defining characteristics of M dwarf photospheres
is the formation of molecules.  And for Si, the dominant molecule
will be SiO \citep{tt73,phh99}.  The other low-FIP
elements that we are concerned with, Mg and Fe, are less inclined to
form molecules, and the molecules that they do form (e.g., MgH and
FeH) are more easily dissociated than SiO, which has a relatively high
dissociation energy of $8.18\pm 0.17$~eV \citep{rrr98}.  We
propose that the robustness of SiO leads to Si behaving
more like a high-FIP element for M dwarfs, leading to the high Si/Fe
ratios seen in Figure~7(d).  Exploring this hypothesis further
would require more detailed modeling than we can provide here.

     Finally, the coronal Ne abundances are worthy of discussion,
due to their relevance for establishing the true cosmic
abundance of Ne.  Solar photospheric measurements are often used as
reference ``cosmic abundances'' throughout astronomy, but there are no
photospheric Ne lines, so the solar Ne abundance is instead
estimated from transition region and coronal lines \citep{jts05,pry05}.
Considering that Ne is one of the most
abundant elements in the universe, establishing the proper cosmic
abundance of Ne has broad ramifications.  The Ne abundance is often
measured relative to O, another high-FIP element.  The reference solar
photospheric abundances used here, from \citet{ma09}, assume
${\rm Ne/O}=0.17\pm0.05$.  However, stellar coronal measurements tend
to be much higher than this, with \citet{jjd05} finding a
weighted mean of ${\rm Ne/O}=0.41$ for a sample of 23 stars, with none
of the stellar measurements as low as the solar one.

\begin{figure}[t]
\plotfiddle{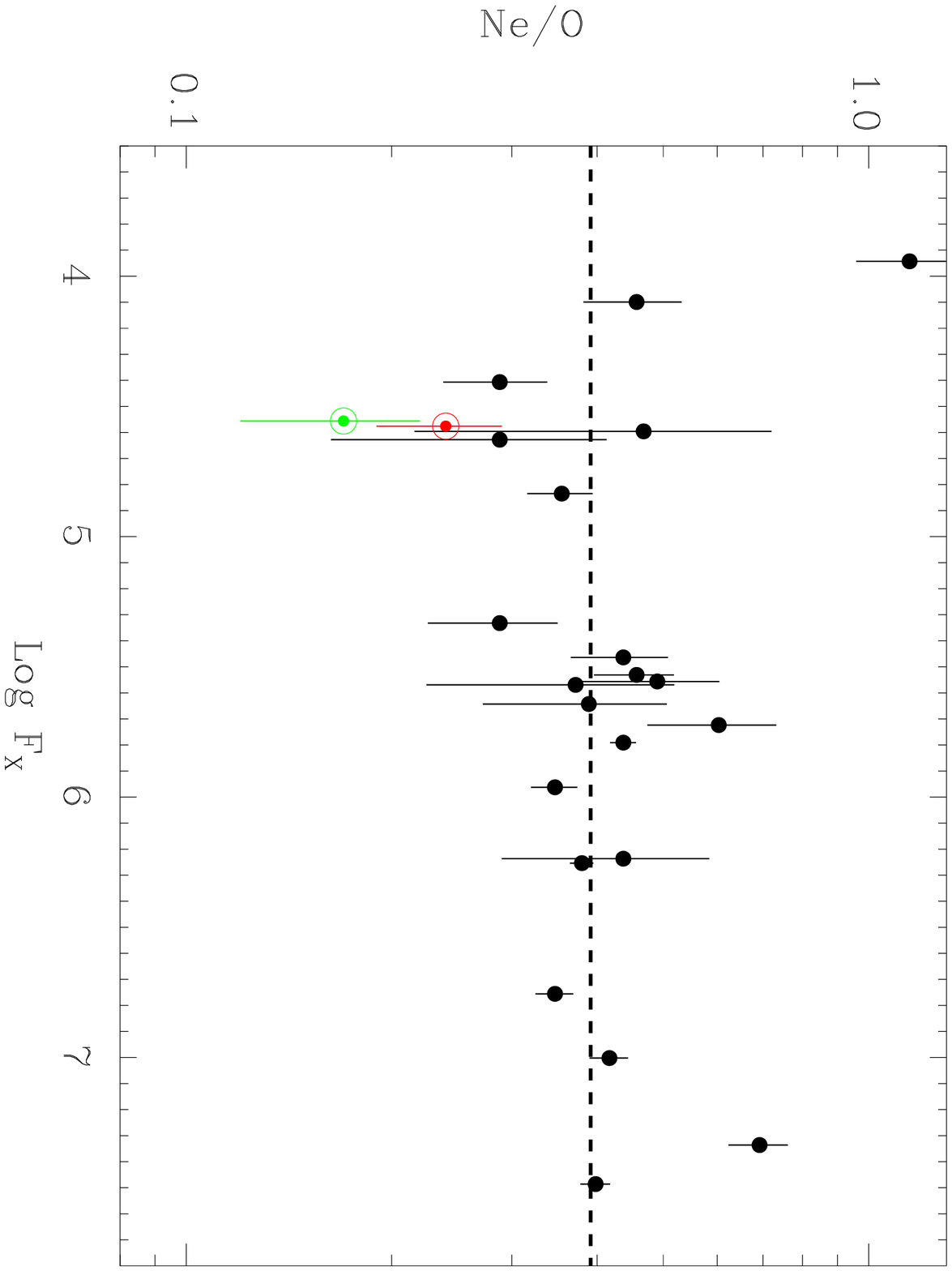}{3.5in}{90}{55}{55}{210}{-40}
\caption{Coronal Ne/O ratios of main sequence stars observed by
  {\em Chandra}/LETGS versus X-ray surface flux.  The
  dashed line is the weighted mean.  Two values of the solar Ne/O are
  shown for comparison, from \citet{ma09} in green
  and \citet{pry18} in red.}
\end{figure}
     This raises two questions:  1. Why are the solar and stellar
Ne/O measurements in disagreement, and 2. Which value should be used
to define the reference cosmic Ne abundance?  As for the first
question, \citet{jr08} found evidence for lower stellar Ne/O
ratios for less active stars, and argued that the low solar value may
not be as inconsistent as the \citet{jjd05} sample suggests,
since the sample is dominated by stars far more active than the Sun.
In Figure~8, we look for an activity correlation for Ne/O within our
sample of main sequence stars.  The Ne/O values in the figure are
from Table~4, with uncertainties estimated from the standard
deviation of the Ne/O values computed during PINTofALE's numerous
Monte Carlo trials.  Excluding the anomalously high value for
$\alpha$~Cen~A(lo), we find a weighted mean of ${\rm Ne/O}=0.39$, in
good agreement with \citet{jjd05}.  However, unlike \citet{jr08},
we see no evidence for any activity dependence,
and like the \citet{jjd05} sample, not one of our measurements
is as low as the solar one.

     Nevertheless, the solar/stellar discrepancy has still
improved somewhat, thanks mostly to the recent upward revision
of the solar Ne/O ratio by \citet{pry18} to ${\rm Ne/O}=0.24\pm 0.05$.
(see Figure~8).  This measurement is from the quiet Sun
transition region.  \citet{el15} find that Ne/O can be
variable in the solar corona, with a high value of
${\rm Ne/O}=0.25\pm 0.05$ corresponding to the
lowest activity corona, consistent with \citet{pry18}.
The most relevant comparison stars for the Sun
in our sample are $\alpha$~Cen~A and B.  Our four measurements
for those two stars are ${\rm Ne/O}=1.15\pm 0.19$, $0.46\pm 0.08$,
$0.29\pm 0.05$, and $0.36\pm 0.04$ for $\alpha$~Cen~A(lo),
$\alpha$~Cen~A(hi), $\alpha$~Cen~B(lo), and $\alpha$~Cen~B(hi), 
respectively.  The very high $\alpha$~Cen~A(lo) value is clearly
anomalous, and we disregard it when computing the weighted mean
quoted above.  The $\alpha$~Cen~A(lo) spectrum is the
only one in our sample where Ne~IX $\lambda$13.5 is undetected and
O~VIII $\lambda$19.0 is just barely detected.  The other $\alpha$~Cen
Ne/O values, however, are reasonably consistent with those of the more
active stars in our sample.  Note that \citet{cl06}
measured ${\rm Ne/O}=0.28$ from XMM spectra of $\alpha$~Cen~AB
combined, which will be dominated by the brighter $\alpha$~Cen~B.
This is only a little lower than our $\alpha$~Cen~B measurements.

     It remains an open question as to whether the cosmic Ne
abundance is best assumed to be the new ${\rm Ne/O}=0.24\pm 0.05$
solar value from \citet{pry18}, or our new stellar average of
${\rm Ne/O}=0.39$.  The solar value has the advantage of being
from the transition region, where fractionation effects are believed
to be less pronounced than in the corona.  The stellar measurement
has the advantage of coming from multiple sources.

\section{The Disconcerting Absolute Abundance Measurements}

     The term ``FIP effect'' is often used to describe the coronal
abundance anomalies seen in the solar corona.  But the $F_{bias}$ and
[Fe/Fe$_*$] quantities in Figures~7(b-c) represent two fundamentally
different ways of thinking about the ``FIP effect,'' the former
involving relative abundances and the latter involving absolute
abundances.  This can lead to significant confusion with regards
to terminology.  If it is said that a star has a solar-like FIP
effect, does this mean low-FIP elements are enhanced relative to
high-FIP elements, or does it mean that the absolute abundances of
low-FIP elements are enhanced?

     There is no distinction if high-FIP
elements are not fractionated relative to H, which is after all
a high-FIP element.  The dashed line in Figure~7(c) shows explicitly
what the [Fe/Fe$_*$] curve should look like if high-FIP elements
were rigidly tied to H, based on equation~(5).  There is significant
discrepancy from the observed relation.  The comparison
suggests that high-FIP coronal abundances are roughly photospheric
for A and M stars, but are depleted for FGK stars.  The Sun provides
some evidence that high-FIP depletion is at least possible, with
He in the slow solar wind being depleted by about a factor of two
\citep{rvs95}.

     Figure~7(b) shows clearly that for solar-like G stars, low-FIP
elements are enhanced relative to high-FIP elements, and that the
stellar measurements are nicely consistent with those of the solar
corona.  In contrast, the linear fit in Figure~7(c) seems to imply
that the absolute abundances of Fe in G star coronae are little
different from photospheric, meaning that the relative abundance
effect in Figure~7(b) is actually due primarily to depletion of
high-FIP elements rather than an enhancement of low-FIP elements, in
contrast to what is assumed to be the case for the Sun.
In short, it is not clear that the G dwarf absolute abundance
measurements are consistent with the solar coronal measurements.
The solar data point seems to be uncomfortably high in Figure~7(c),
higher than the other G stars, and higher than all other stars except
for Altair.  The problem is even worse if the \citet{jts12}
measurement of solar [Fe/Fe$_*$] is replaced
with the older but more canonical [Fe/Fe$_*$]=+0.6 value of
\citet{uf00}.  This replacement would make the solar point
in Figure~7(c) even more inconsistent with the stellar relation.
So are we to say that early G stars have solar-like FIP effects or
not?  In terms of relative abundances (e.g., Figure~7b) they
definitely do, but in terms of absolute abundances (e.g., Figure~7c)
they may not.

     Both the significant scatter seen in Figure~7(c) and the
questionable consistency between the solar and stellar measurements
cast doubt on the accuracy of the absolute abundance measurements,
which rely on the line-to-continuum ratio analysis.  Possible pitfalls
in this analysis have been discussed by \citet{jjd98} and
\citet{mg04a}.  A particularly important one involves
the possibility that extensive blends of weak emission lines that
are not in CHIANTI might be mistaken for the continuum.  Is it possible
that our continuum estimates in Figure~5 (blue lines) are too high
due to the presence of a plethora of unknown, highly blended, weak
emission lines in the $25-40$~\AA\ region?  This would
lead to underestimates of [Fe/Fe$_*$], and if ubiquitous could push
the stellar values upwards to be more consistent with the currently
accepted solar value.
\citet{el13} provide an example of how numerous
weak lines added in CHIANTI v7.1 significantly change and
improve the appearance of synthetic spectra from 80--120~\AA,
when compared with CHIANTI v7.0.  A similar change in the
$25-40$~\AA\ region would have a dramatic effect on our
line-to-continuum analysis.

     It is worth noting that solar studies also have a history of
ambiguity with regards to the issue of whether the solar coronal FIP
effect is one of low-FIP enhancement or high-FIP depletion
\citep{uf03}.  Past claims of high-FIP depletions include
\citet{njv81}, \citet{af95}, and \citet{jcr97}.
Current preference for the low-FIP enhancement
interpretation comes in part from direct particle measurements of the
slow solar wind, which are more suggestive of low-FIP enhancement and
photospheric high-FIP; except for the aforementioned high-FIP He, with
its factor of two depletion \citep{rvs00}.
Radio observations of the Sun also seem to provide
support for a roughly factor of four enhacement of low-FIP elements in
the corona \citep{smw00,sjs15}.

     Direct comparison of the Sun with the most solar-like stars in our
smaple, $\alpha$~Cen A and B, provides further cause for unease
with regards to the absolute abundance measurements.
In Figure~4 the EM distributions of $\alpha$~Cen A and B, are
compared with solar distributions from \citep{sjs17},
from full-disk SDO/EVE spectra.  The three solar distributions,
from Figure~9 of \citet{sjs17}, represent a range of
activity states for the solar corona.  Significant variation
in $EM_V$ is observed only for $\log T>6.1$.  This is very
consistent with what we find when comparing the ``(lo)'' and ``(hi)''
EM distributions for $\alpha$~Cen~A and B, where we see little
variation for $\log T<6.1$.

     Although it is $\alpha$~Cen~A that is generally considered the
true Sun-like star of the $\alpha$~Cen binary, due to its identical
G2~V spectral type, it is $\alpha$~Cen~B that seems
most similar to the Sun in terms of coronal properties.  Not
only are the $\log F_X$ values measured for $\alpha$~Cen~B the most
solar-like, the shape of the $\alpha$~Cen~B EM distribution is also
the most similar to the Sun.  In contrast, the $\alpha$~Cen~A
corona is cooler and has significantly lower $F_X$.
Despite the coronal similarities, the magnitudes of $EM_V$ are
somewhat lower for the solar distributions than for $\alpha$~Cen~B.
This is mostly due to the different [Fe/Fe$_*$] assumed in
normalizing the $EM_V$ curves.  \citet{sjs17} simply assume
${\rm [Fe/Fe_*]}=+0.6$ \citep{sjs15}, while for $\alpha$~Cen~B
we have measured identical ${\rm [Fe/Fe_*]}=-0.61$ values from both the
``(lo)'' and ``(hi)'' spectra.  The higher reference photospheric Fe
abundance for $\alpha$~Cen~B (${\rm [Fe_*/Fe_{\odot}]}=0.27$ from Table~1)
moderates this discrepancy somewhat, but the coronal Fe abundance
difference still adds up to $0.94$~dex.  The $\alpha$~Cen~B panel in
Figure~4 shows the improved agreement between the intermediate solar
$EM_V$ curve and the $\alpha$~Cen~B(hi) distribution when the solar
$EM_V$ curve is renormalized to assume the $\alpha$~Cen~B coronal Fe
abundance.  Could it really be true that the Sun and $\alpha$~Cen~B
are coronally so similar in terms of X-ray flux and
temperature distribution, but still exhibit radically different
absolute abundance behavior, with the Sun having a substantial
enhancement of Fe and $\alpha$~Cen~B having a dramatic
depletion?

     Another coronal abundance comparison that can be made
between the Sun and $\alpha$~Cen~AB concerns the issue of activity
cycle variability.  \citet{dhb17} present evidence for a solar
cycle variation of FIP bias, with a stronger FIP effect at solar
maximum.  However, for both $\alpha$~Cen~A and B, we find no
significant difference in the $F_{bias}$ or [Fe/Fe$_*$] measurements
from the ``(lo)'' and ``(hi)'' spectra.  This conclusion is evident even
without an EM analysis, as shown in Figure~3(d).  Thus, we conclude
that $\alpha$~Cen~AB coronal abundances do not vary significantly
during the stellar activity cycles \citep{tra14}.  The tightness of the
FBST relation in Figure~7(a-b) by itself might imply little time
variation of $F_{bias}$, considering the stars will have been observed
at different points in their various activity cycles.  However, most
of these stars are significantly more active than the Sun, and such
stars tend to have more irregular activity cycles \citep{ko16,rrr18}.

\section{Emission Measure Variation with Activity}

\begin{figure}[t]
\plotfiddle{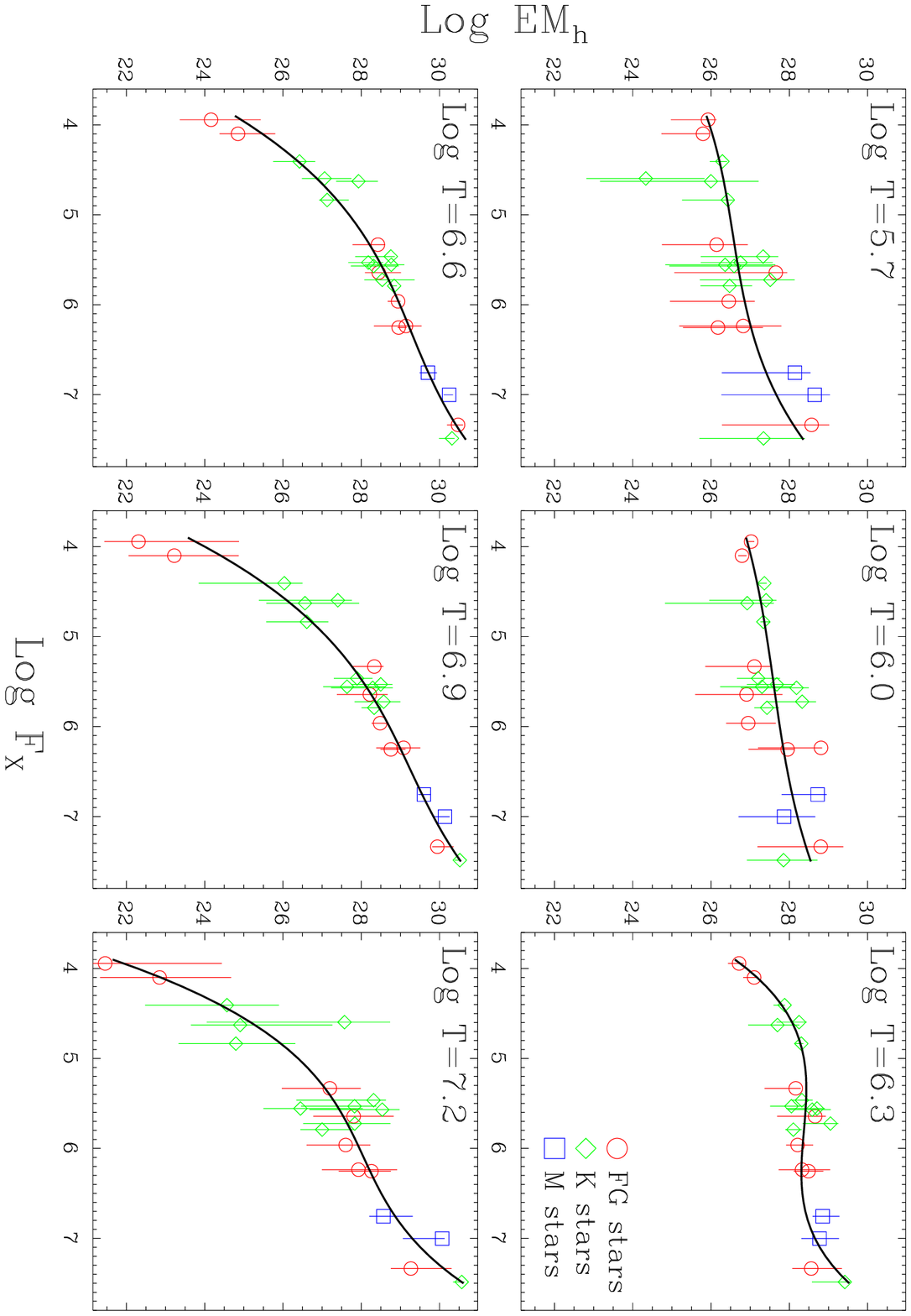}{4.7in}{90}{80}{80}{315}{-70}
\caption{Emission measure at a given temperature versus stellar
  X-ray surface flux, based on the EM distributions in Figure~4.  The
  relations are shown for six different temperature bins.  The data
  are fitted with third order polynomials.}
\end{figure}
     We now use the sample of EM distributions in Figure~4 to assess
how the EMs of main sequence stars vary with
stellar activity.  The Figure~4 distributions are provided in
temperature bins with widths of 0.1~dex in $\log T$.  For each $\log T$
bin, we can plot the measured EM values at that temperature versus $F_X$
for our sample of stars.  Six of these plots are shown in Figure~9, for
six different temperatures.  We have converted from $EM_V$ to $EM_h$
to correct for the different radii of our stars.  For each temperature
bin, we find a reasonably smooth and consistent variation of $EM_h$
with $F_X$, with no evidence for any substantial spectral type
variation.  The M, K, and FG dwarfs seem to be consistent with
each other.  We fit third-order polynomials to the data points.
Increases in $EM_h$ with $F_X$ are seen at all temperatures, but the
relations are relatively flat for $\log T\leq 6.3$.  The slope of
the relation increases greatly for $\log T>6.3$.  It is
remarkable that at $\log T=6.3$, $EM_h$ for the extremely active
EK~Dra is not much higher than for the comparatively inactive
$\alpha$~Cen~B(hi), but at $\log T=6.6$ the difference
balloons to over three orders of magnitude.

\begin{figure}[t]
\plotfiddle{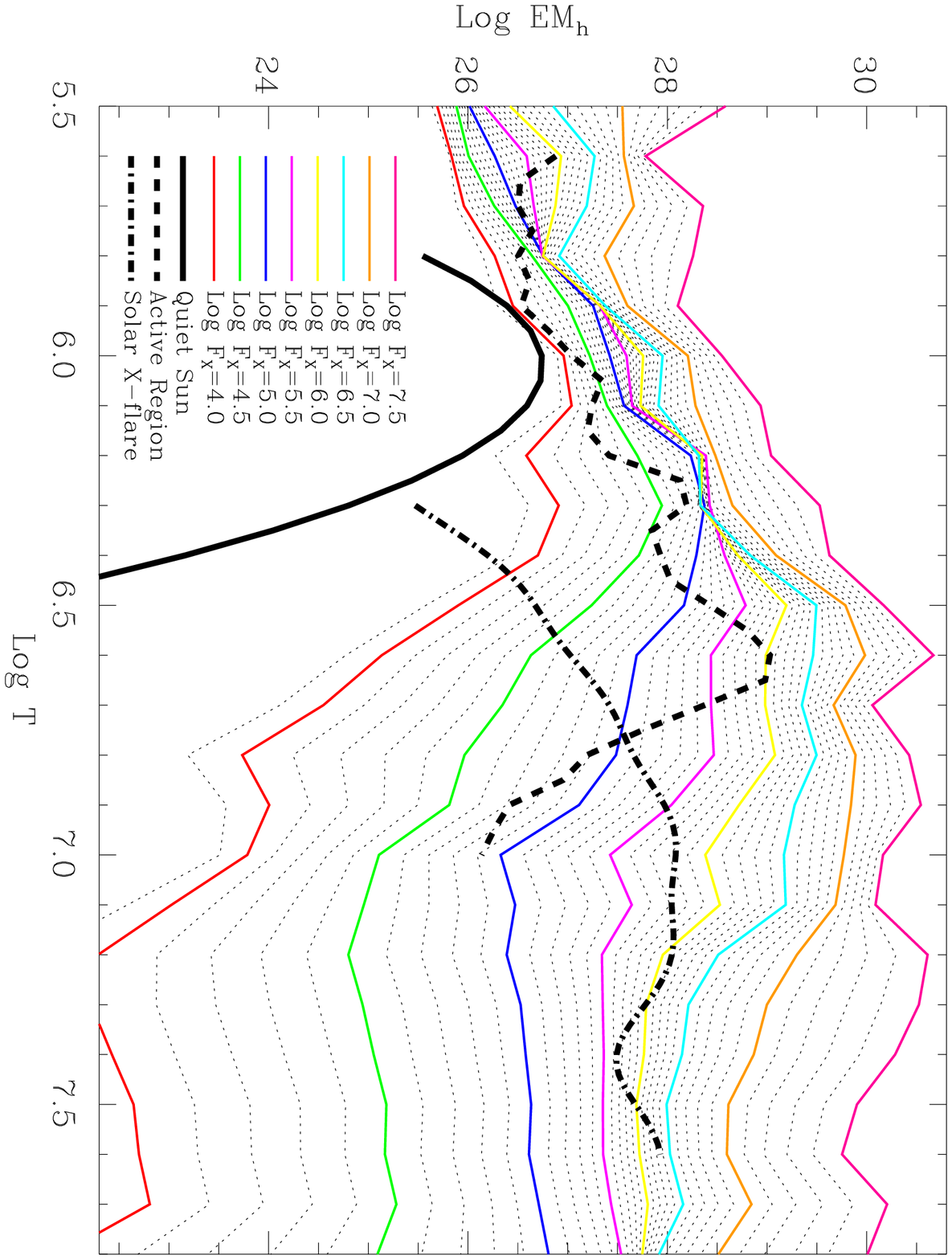}{4.5in}{90}{70}{70}{270}{-50}
\caption{Main sequence star emission measure distributions as a function
  of $F_X$, based on relations such as those in Figure~9.  These are
  compared with an average solar active region distribution from
  \citet{hpw12}, a quiet Sun distribution from \citet{sk12},
  and an average solar flare distribution from \citet{hpw14}.
  The first two are from spatially resolved data,
  so these are true line-of-sight column emission measures, while the
  flare distribution is from disk-integrated data, so the $EM_h$
  values are derived assuming the emission is uniformly spread over
  the visible surface.}
\end{figure}
\begin{table}
\scriptsize
\begin{center}
\caption{Emission Measure Variation with Stellar Activity$^a$}
\begin{tabular}{ccccccccc} \hline \hline
 & \multicolumn{8}{c}{$\log F_X$ (ergs cm$^{-2}$ s$^{-1}$)} \\
\cline{2-9}
$\log T$ & 4.0 & 4.5 & 5.0 & 5.5 & 6.0 & 6.5 & 7.0 & 7.5 \\
\hline
5.5  & 25.69 & 25.88 & 26.02 & 26.17 & 26.42 & 26.85 & 27.55 & 28.58 \\
5.6  & 25.83 & 26.01 & 26.27 & 26.59 & 26.93 & 27.27 & 27.56 & 27.78 \\
5.7  & 25.96 & 26.26 & 26.48 & 26.66 & 26.88 & 27.19 & 27.66 & 28.36 \\
5.8  & 26.27 & 26.64 & 26.76 & 26.75 & 26.75 & 26.92 & 27.37 & 28.26 \\
5.9  & 26.45 & 27.00 & 27.26 & 27.32 & 27.33 & 27.38 & 27.60 & 28.11 \\
6.0  & 26.96 & 27.22 & 27.42 & 27.59 & 27.76 & 27.95 & 28.21 & 28.55 \\
6.1  & 27.04 & 27.39 & 27.57 & 27.65 & 27.73 & 27.91 & 28.28 & 28.94 \\
6.2  & 26.59 & 27.70 & 28.23 & 28.39 & 28.35 & 28.32 & 28.48 & 29.04 \\
6.3  & 26.91 & 27.95 & 28.37 & 28.42 & 28.33 & 28.33 & 28.65 & 29.53 \\
6.4  & 26.70 & 27.71 & 28.29 & 28.57 & 28.71 & 28.83 & 29.09 & 29.63 \\
6.5  & 25.90 & 27.24 & 28.17 & 28.79 & 29.19 & 29.50 & 29.79 & 30.17 \\
6.6  & 25.14 & 26.63 & 27.69 & 28.44 & 28.99 & 29.46 & 29.98 & 30.67 \\
6.7  & 24.55 & 26.34 & 27.60 & 28.44 & 28.98 & 29.35 & 29.67 & 30.06 \\
6.8  & 23.73 & 25.96 & 27.48 & 28.47 & 29.08 & 29.50 & 29.89 & 30.43 \\
6.9  & 24.01 & 25.81 & 27.11 & 28.04 & 28.72 & 29.28 & 29.84 & 30.54 \\
7.0  & 23.79 & 25.11 & 26.33 & 27.43 & 28.38 & 29.17 & 29.77 & 30.16 \\
7.1  & 23.03 & 24.95 & 26.47 & 27.64 & 28.53 & 29.19 & 29.69 & 30.09 \\
7.2  & 22.29 & 24.80 & 26.39 & 27.34 & 27.95 & 28.51 & 29.30 & 30.61 \\
7.3  & 22.22 & 24.94 & 26.53 & 27.35 & 27.79 & 28.21 & 29.00 & 30.52 \\
7.4  & 22.43 & 25.05 & 26.58 & 27.36 & 27.77 & 28.15 & 28.87 & 30.29 \\
7.5  & 22.64 & 25.18 & 26.63 & 27.35 & 27.69 & 27.99 & 28.61 & 29.90 \\
7.6  & 22.70 & 25.17 & 26.61 & 27.36 & 27.72 & 28.03 & 28.60 & 29.75 \\
7.7  & 22.81 & 25.28 & 26.71 & 27.44 & 27.80 & 28.16 & 28.84 & 30.20 \\
7.8  & 21.91 & 25.09 & 26.81 & 27.54 & 27.75 & 27.92 & 28.51 & 30.01 \\
\hline
\end{tabular}
\end{center}
\tablecomments{$^a$Emission measures are $\log EM_h$ values (units cm$^{-5}$)
  from Figure~10.}
\normalsize
\end{table}
     The polynomial fits to the $EM_h(T)$ versus $F_X$ relations can be
used to define an average EM distribution as a function of $F_X$.
This is done in Figure~10, which shows how a main sequence star
$EM_h$ distribution varies with $F_X$ for $\log F_X=3.9-7.5$.
These curves are also provided in Table~7.  This information can be
used to estimate an EM distribution for any main sequence star that only
has a broadband X-ray flux measurement.  It can also be used as the
basis for future theoretical coronal heating models seeking to describe
how heating changes with increasing stellar activity.  If mean
coronal temperatures are computed from these curves as prescribed
by \citet{cpj15}, we can reproduce their power law
relation between $T_{cor}$ and $F_X$.  Specifically, with $T_{cor}$ in
MK units we find $T_{cor}=0.16 F_X^{0.24}$, compared with
Johnstone \& G\"{u}del's $T_{cor}=0.11 F_X^{0.26}$.

     The stellar EM distributions in Figure~10 are compared with
three solar distributions, representing the quiet Sun (QS), solar
active region (AR), and flaring (FL) Sun.  The QS distribution is
based on spectra from the EUV Imaging Spectrometer (EIS) on
{\em Hinode} \citep{sk12}.  The AR distribution is an
average of 15 ARs studied by \citet{hpw12} using EIS and
SDO/AIA data.  Finally, the FL distribution is an average distribution
from the peaks of 21 strong flares studied by \citet{hpw14} using
SDO/EVE spectra.

\begin{figure}
\plotfiddle{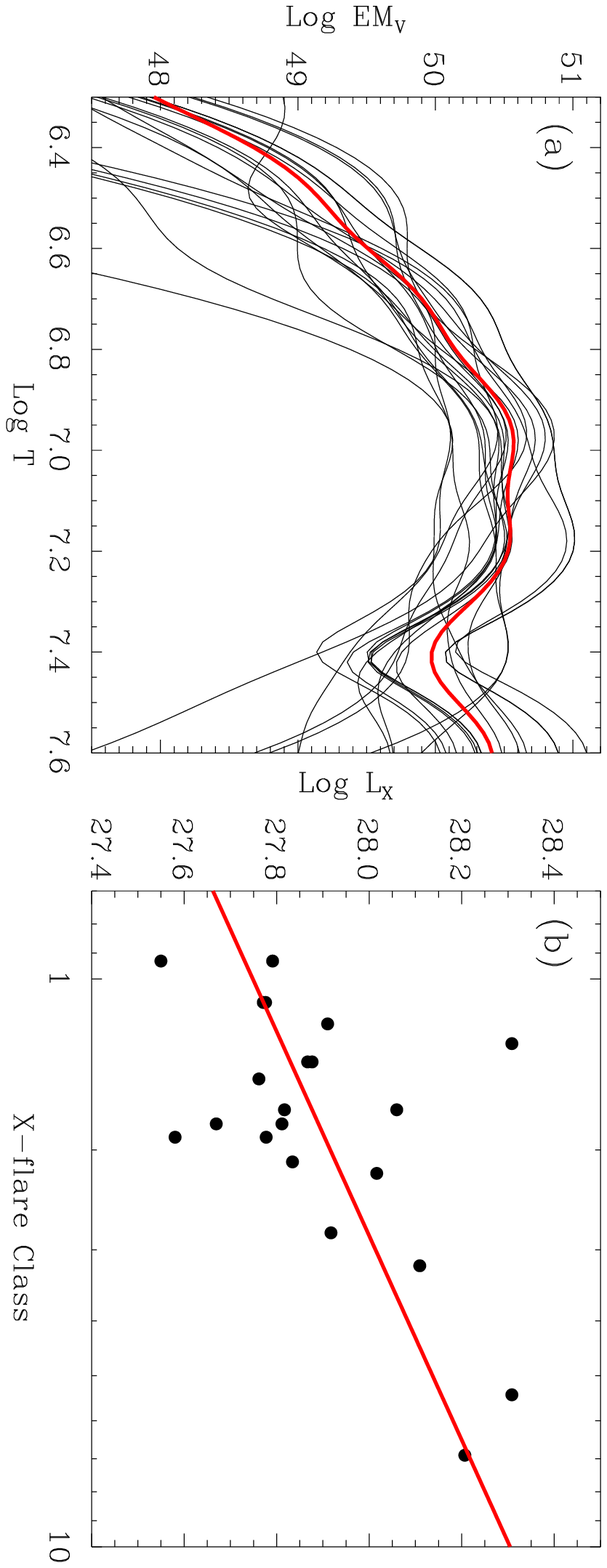}{2.9in}{90}{80}{80}{310}{-240}
\caption{(a) Solar flare emission measure distributions from \citet{hpw14},
  and their mean in red.  (b) Solar flare X-ray luminosities versus GOES
  X-flare class, with a linear fit.}
\end{figure}
     In solar physics, decades of flare monitoring with the
Geostationary Operational Environmental Satellite (GOES) spacecraft
have led to the widespread use of the GOES system's classification of
flare strength, with the A, B, C, M, and X classes representing
increasing decades of flare luminosity in the $1-8$~\AA\ bandpass.
The 21 flares used to define the FL distribution range from M9.3 to
X6.9.  The individual flare EMs are shown explicitly in Figure~11(a),
where we focus only on the 2-minute interval of maximum inegrated EM.
Besides using them to compute the average FL EM for Figure~10, we also
compute synthetic spectra from them, and from those spectra we
estimate X-ray luminosities in the $5-120$~\AA\ bandpass most familiar
to stellar astronomers.  In Figure~11(b), these X-ray luminosities are
plotted versus the GOES X-flare classification, $X_f$ (e.g., for
an X5.2 flare, $X_f=5.2$; for an M7.5 flare, $X_f=0.75$).  Not
surprisingly, $\log L_X$ increases with flare strength, with the fit
to the data suggesting $\log L_X=0.56\log X_f+27.75$.  A typical
X-flare has $\log L_X\approx 28.0$, which is about five times the
Sun's quiescent X-ray luminosity, but is less than the quiescent
emission from most of our stars (see Table~1).

     There is an important distinction between the QS and AR
distributions and the FL distribution in Figure~10.  The former are
from spatially resolved observations, so the $EM_h$ values are
line-of-sight column emission measures.  Thus, these curves are
indicative of the actual intensities of QS and AR regions.  In
contrast, the FL measurements are from disk-integrated SDO/EVE
spectra, so $EM_h(T)$ is computed in the same manner as for the stars,
namely by computing $EM_V$ and then converting to a column $EM_h$ by
dividing by the visible surface area of the star.  In order to scale
the FL $EM_h$ curve to indicate the actual flare intensity, it is
necessary to divide the curve by the fractional surface coverage of a
flare.  For the 21 flares from \citet{hpw14}, we estimate an average
surface coverage of 0.3\%.  This corresponds to increasing the FL
curve in Figure~10 by 2.5~dex.

     We can compute synthetic spectra and X-ray luminosities
for each of the solar EM distributions in Figure~10, and for
the \citet{sjs17} full-disk solar EM distributions in Figure~4.  We
find $\log L_X=26.40$, 28.66, and 27.94 for the QS, AR, and FL
distributions, respectively, while the \citet{sjs17} distributions
imply a range of $\log L_X=26.86-27.40$.  The QS value represents an
estimate of the minimum possible X-ray luminosity for a solar-like
G star.  To generalize to main sequence stars with different radii,
this corresponds to a minimum surface flux of $\log F_X=3.61$.
This is only about a factor of two lower than the lowest
activity spectrum in our sample, that of $\alpha$~Cen~A(lo).
We are not aware of any main sequence stars with detected X-ray fluxes
below this limit \citep[e.g.,][]{js04}.  In principle, lower X-ray
luminosities might be expected for very low metallicity stars, due to
weak emission lines.  However, the models of \citet{tks18} suggest
that such stars may end up brighter in X-rays due to much higher
coronal densities.

     \citet{pgj17} recently found a very low upper limit for
the X-ray luminosity of 16~Cyg~B (G3~V), $\log L_X<25.5$, which is
much lower than the minimum QS level of $\log L_X=26.4$ just quoted in
the previous paragraph.  However, the 16~Cyg observations are
{\em Chandra} ACIS-I data, which covers an energy range of $0.3-2.5$~keV,
compared to the more traditional $0.1-2.4$~keV {\em ROSAT}/PSPC range
that we are using.  These ranges seem similar, but the similarity is
illusory for very inactive stars, as an inactive star will emit far
more flux in the $0.1-0.3$~keV bandpass than it will at $>0.3$~keV,
due to the very low coronal temperatures.  At $>0.3$~keV, an extremely
inactive star's emission will be mostly from the O~VII triplet at
22~\AA\ and the C~VI line at 33.7~\AA.  For an energy range of
$0.3-2.5$~keV we find $\log L_X=25.37$ and $\log F_X=2.59$ for the QS
distribution, a full order of magnitude lower than the $0.1-2.4$~keV
values quoted above.  This $0.3-2.5$~keV QS luminosity is {\em not}
inconsistent with the very low 16~Cyg~B limit of \citet{pgj17}.  The
luminosities inferred from the \citet{sjs17} distributions decrease
to $\log L_X=26.15-26.96$ in the
$0.3-2.5$~keV range, more consistent with the $\alpha$~Cen~AB
luminosities quoted by \citet{tra14}, which are for $0.2-2$~keV.

     It is impressive that the stellar $EM_h$ distributions in
Figure~10 end up matching the solar AR distribution so well for $\log
T<6.6$, in terms of both magnitude and slope of the distribution.
This is true despite the solar and stellar analyses relying on
completely different sets of emission lines measured from completely
different wavelength regions.
Between $\log F_X=5.0$ and $\log F_X=6.5$, the $EM_h$ curves basically
lie right on top of each other for $\log T<6.4$, suggestive of a
saturation effect.  This apparent saturation at close to the $EM_h$
level of the solar AR distribution strongly supports the idea first
suggested by \citet{jjd00} that the EM distributions of
intermediate activity stars like $\epsilon$~Eri and $\xi$~Boo~A are
nicely explained by the stellar surfaces being completely filled with
solar-like ARs.  The $EM_h$ values are very linear for $\log T<6.6$,
so we can assume $EM_h(T)\propto T^{\beta}$ and measure the slope,
$\beta$.  For the stellar EMs with $\log F_X=5.5-6.5$ we find a mean
slope of $\beta=2.50$, in excellent agreement with the solar AR slope
of $\beta=2.41$.  The EM seems to start increasing again for $\log
F_X>7.0$, but there are only a few stars in this high activity regime,
so the reality of this increase is questionable.

     For the solar AR distribution, $EM_h$ drops precipitously
for $\log T>6.6$.  It is therefore clear that while solar-like
ARs may explain the stellar $EM_h$ for $\log T<6.6$, they cannot
account for the high temperature $EM_h$ at $\log T>6.6$, which
becomes quite strong for stars with $\log F_X>5.5$.  For this
emission, the only solar analog would be solar flares, as
represented by the FL distribution in Figure~10.  Simply adding
the AR and FL distributions represents a decent approximation of the
stellar distributions corresponding to $\log F_X\sim 6.0$, suggesting
that such stars are completely covered with solar-like ARs, but also
have the equivalent of a single X-class flare's worth of high
temperature emission occurring at all times.  Recalling the 0.3\%
surface coverage of individual solar flares noted above, it is worth
noting that completely covering the surface of a star with X-class
flares would lead to a reasonably accurate representation of the EM
distributions of the highest activity stars in our sample, with
$\log F_X\sim 7.5$.

     Coronal density measurements provide support for the
$\log T\sim 7.0$ emission being fundamentally different from the
cooler AR-like emission at $\log T\sim 6.5$.  Analyses of
density-sensitive line ratios generally find electron densities (in
units of cm$^{-3}$) of $\log n_e\sim 10$ for $\log T<6.6$, consistent
with typical solar AR densities.  However, much higher densities of
$\log n_e\sim 12-13$ are generally observed at $\log T\sim 7$
\citep{am04,pt04,rao06,cl08}.

     It is worth looking for abundance signatures that might also
indicate a distinction between the $\log T\sim 6.5$ and
$\log T\sim 7.0$ plasma.  Unlike the quiescent solar corona,
solar flares generally do not exhibit a FIP effect.  This is
demonstrably the case for the 21 SDO/EVE solar flares we have used to
define the FL distribution in Figure~10, which have roughly
photospheric abundances \citep{hpw14}.  However, it is very hard for
us to look for a FIP effect specifically at $\log T\sim 7.0$, because
the only high-FIP line we have that is at least nominally formed at
$\log T>6.6$ is Ne~X.  This is illustrated best by Figure~3, which does
not suggest dramatic differences in FIP bias at Ne~X temperatures
compared to lower temperatures.  However, even Ne~X is problematic,
because the contribution function for this species is quite broad
in temperature, meaning that if $EM_V$ is higher at $\log T=6.5$ than at
the nominal peak at $\log T\approx 6.9$ by as little as a factor of 3 or
4, then much of the Ne~X emission will actually be coming from
$\log T\approx 6.5$.  Most of our stellar EM distributions do in fact
peak near $\log T=6.5$.


\section{Exoplanet Effects on Coronal Abundances}

     Ever since the first detection of ``hot Jupiters'' orbiting
very close to their host stars, there has been the question of what
effects this close proximity will have on both the planet and star.
There is no doubt that the planet's characteristics will be greatly
affected by being so close to the star.  Its temperature will be
blisteringly hot, and its atmosphere will be exposed to extremely high
fluxes of both radiation and stellar wind.

     More surprising is the notion that the close-in planet
might actually affect the star as well, particularly its atmospheric
activity.  This is possible in principle through tidal or
magnetospheric interactions \citep{mc00,oc09}.
Several cases have been reported of chromospheric variability with
periods commensurate with that of the planet \citep{es08,gahw08},
but these must be balanced against more numerous
examples of nondetections \citep{es05,kp11a,bpm12,gs13}.  There have
been claims of enhanced activity in planetary host stars in general
\citep{vlk08,cas10}, but such correlations may just
be a spurious consequence of selection effects \citep{kp11b,blcm11}.

     Focusing on particularly close-in, massive exoplanet cases where
tidal effects are most likely to be present, \citet{kp14}
find at least a couple cases where an exoplanet host's activity level
is anomalously high compared to a distant stellar companion, suggesting
that perhaps tidal effect from the exoplanet are keeping the stellar
host spun-up and more active than the companion star.  These two
cases are HD~189733A and CoRoT-2A, which host hot Jupiters with
$P_{orb}=2.22$~day and $P_{orb}=1.74$~day, respectively.  However,
there are also cases where the exoplanet seems to be {\em inhibiting}
stellar activity.  The WASP-18 system seems to be
such a case, with $P_{orb}=0.96$~days and with stellar X-ray and UV
emission both anomalously low \citep{ip14b,lf18}.
For systems such as this, tidal effects
of the planet may actually be adversely affecting the stellar dynamo.

\begin{figure}[t]
\plotfiddle{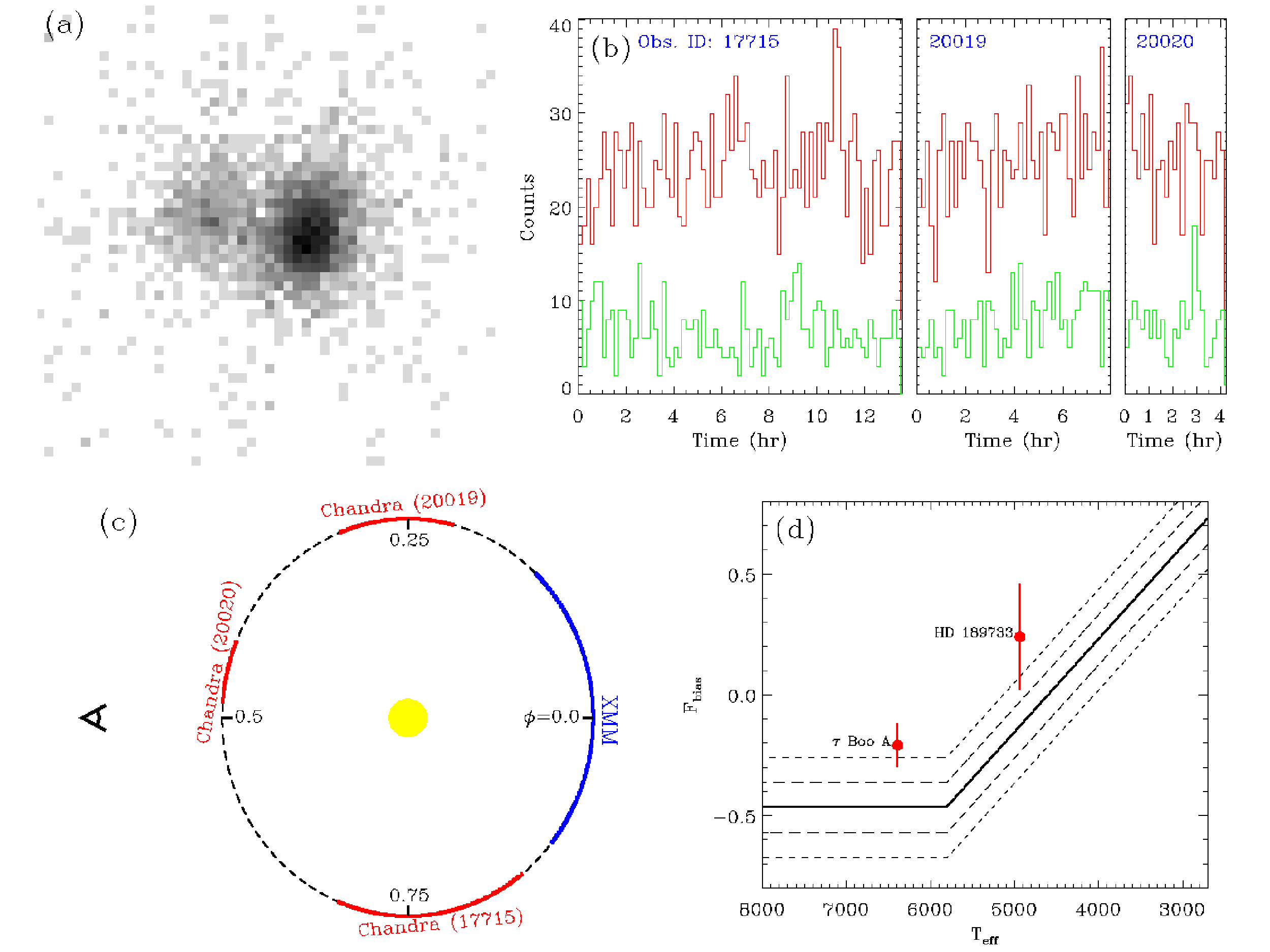}{4.6in}{0}{65}{65}{-230}{0}
\caption{(a) Zeroth-order {\em Chandra}/LETGS image of the $\tau$~Boo~AB
  binary (F7~V+M2~V), from observation ID 17715. (b) Light curves of
  $\tau$~Boo~A (red) and B (green), extracted from the zeroth-order
  images during the three separate LETGS exposures. (c) Schematic
  illustration of the orbital phases of $\tau$~Boo~A's exoplanet
  sampled by the {\em Chandra} exposures from 2017, and the XMM
  exposure from 2003. (d) Reproduction of the $F_{bias}$ versus
  temperature relation from Figure~7(b), but showing only the data
  points of two exoplanet host stars, which both lie above the
  relation.}
\end{figure}
     One of the examples of a planet affecting a stellar atmosphere
involves $\tau$~Boo~A, which is one of the stars in our sample.
As noted in Section~2, $\tau$~Boo~A has an M2~V stellar companion,
$\tau$~Boo~B.  The binary is shown explicitly in the zeroth-order
LETGS image in Figure~12(a).  We measure a position angle and
stellar separation of $\theta=79.2\pm 0.7^{\circ}$ and
$\rho=1.545\pm 0.019^{\prime\prime}$, respectively.  According to
the zeroth order image, $\tau$~Boo~B accounts for only about 21\%
of the binary's total X-ray emission.  The planetary companion of
$\tau$~Boo~A, $\tau$~Boo~b, has an orbital period and mass of
$P_{orb}=3.31$ days and $m\sin i=3.9$~M$_J$, respectively
\citep{rpb97}.  The orbital period is about the same as the
stellar rotation period, so this is a case where tidal
synchronization may have taken place \citep{jfd08}.
\citet{es08} and \citet{gahw08} find evidence that $\tau$~Boo~b can
induce an active region on its star that leads the planet by $\sim
70^{\circ}$ in longitude, though the active region may come and go
with time.  The star is also notable for having a remarkably short
activity cycle of 120~days, with detected polarity reversals
\citep{mwm16,mm17,svj18}.  Our goal in this section is to look for evidence
that $\tau$~Boo~A is different from other stars in our sample in
ways that can be interpreted as being an effect of the exoplanet.

     Starting with the EM measurements, there is no indication
that $\tau$~Boo~A's EM's in Figure~9 (at $\log F_X=5.64$) are
systematically discrepant from the other stars.  Thus, we conclude
that there is no evidence that the exoplanet is affecting the coronal
temperature distribution, which looks normal for a star of this
activity level.  Coronal abundances are a different story, however.
We have already noted that $\tau$~Boo~A looks modestly discrepant in
the relative abundance plots in Figure~7(a-b), and more convincingly
discrepant in the absolute abundance plot in Figure~7(c).

     This discrepancy was previously noted by \citet{am11}
and \citet{up15} based on an XMM spectrum from
2003~June~24, though those authors were looking for evidence of
changes in coronal abundance behavior induced by high photospheric
metallicity, rather than changes that might be induced by
$\tau$~Boo~A's exoplanet.  One advantage of the {\em Chandra} spectrum
analyzed here is that the $\tau$~Boo~AB binary is resolved by
{\em Chandra}, unlike XMM, meaning our $\tau$~Boo~A spectrum is
uncontaminated by any emission from the M2~V companion $\tau$~Boo~B.
This is demonstrated explicitly in Figure~12(a).
Nevertheless, we find that the EM distribution and abundances that we
measure for $\tau$~Boo~A agree very well with those reported by
\citet{am11}, suggesting that $\tau$~Boo~B is too faint to have
significantly affected analysis of the XMM data.  In fact, direct
comparison of the {\em Chandra} and XMM spectra shows only very subtle
differences, demonstrating that {\em Chandra}/LETGS and XMM have
impressively consistent flux calibrations, and also that
$\tau$~Boo~A does not vary much between the XMM observation in 2003
and the {\em Chandra} observation in 2017.

     If planets can affect stellar coronae, it would be natural to
suppose that the effects might be greatest on the region of the
star closest to the planet.  Thus, coronal properties might be
expected to be vary with planetary orbital phase.  The {\em Chandra}
observations were fortuitously split into three pieces (see Table~2),
which sample different orbital phases.  Figure~12(b) shows the
light curves of both $\tau$~Boo~A and B measured from these separate
exposures, based on the zeroth-order images.  Using the ephemeris quoted by
\citet{cc07} and \citet{mwm16}, observation IDs
17715, 20019, and 20020 cover orbital phases $\phi=0.69-0.86$,
$\phi=0.21-0.31$, and $\phi=0.44-0.49$, respectively, and
Figure~12(c) schematically shows the orbital geometry corresponding to
these phases.  The $\phi=0$ phase corresponds to the first
conjunction, when the planet is most behind the star from our
perspective, with the planetary orbit tilted by $i=44.5\pm 1.5^{\circ}$
from the plane of the sky \citep{mb12}.  The light curves in
Figure~12(b) and the spectra of the three individual exposures reveal
no clear variation with orbital phase.  Furthermore, the 2003
XMM observations were taken at yet another distinct orbital phase of
$\phi=0.89-1.13$ \citep{am11}, and we have already noted the
lack of significant variation between the
XMM and {\em Chandra}/LETGS spectra.

     We conclude that if the coronal abundance anomalies for
$\tau$~Boo~A apparent in Figure~7 are in fact due to the planet,
the planet must be affecting the corona globally and not locally.
This raises the question as to how an active region on $\tau$~Boo~A
would continue to be affected by a planet that has orbited behind the
star.  However, solar observations suggest that this may be possible
in principle.  In solar ARs, the FIP bias is observed to evolve on
timescales comparable to $\tau$~Boo~b's orbital timescale.  In
particular, newly emerged solar ARs are generally found to have no FIP
effect, and only acquire the normal solar FIP bias after a couple days
\citep{kgw01}.  Therefore, one interpretation of
$\tau$~Boo~A's reduced FIP bias is that the repetitive perturbations
of the exoplanet, with $P_{orb}=3.31$ days, somehow reset the clock
on the stellar ARs, making old ARs behave like young ones.  Exactly
how this might happen is a mystery.  Is it the planet's gravitational
or magnetospheric perturbations that are the most important?

     Demonstrating that $\tau$~Boo~A's exoplanet is truly
the cause of its coronal abundance anomalies requires finding
examples of this effect for other exoplanet host stars.  We take
a first step in this direction using XMM measurements of
O~VIII and Fe~XVII lines from another exoplanet host star,
HD~189733A (K1.5~V), made by \citet{ip14a}.
The transiting planetary companion of this star, HD~189733b, has
an orbital period and semimajor axis of only $P_{orb}=2.22$~days and
$a=0.031$~au, respectively, so this is another very close-in
exoplanet system where tidal effects may be important.  As
mentioned above, HD~189733A is one of the two stars that
\citet{kp14} found to be more active than it
should be compared to its stellar companion's activity level,
suggesting that HD~189733b's presence may be leading to enhanced
activity on HD~189733A.  Using the density-sensitive O~VII lines near
22~\AA, \citet{ip14a} report a potentially anomalous
high coronal density of $n_e=(3-10)\times 10^{10}$~cm$^{-3}$ for
HD~189733A, with 1$\sigma$ error bounds.
The best constraints on these lines for $\tau$~Boo~A would be
from the XMM measurements of \citet{am11} rather than our
LETGS measurements.  These line fluxes are {\em not} suggestive of
high densities, although only a 1$\sigma$ upper limit of
$n_e<3\times 10^{10}$~cm$^{-3}$ can be quoted.

     \citet{bew12} derived a relation between $F_{bias}$ and
the Fe~XVII/O~VIII flux ratio using the Fe~XVII lines at
$15-17$~\AA\ and the O~VIII line at 19.0~\AA.  Based on this relation
and the line fluxes from \citet{ip14a}, we estimate
$F_{bias}=0.16$ for HD~189733A.  However, we have to include
the $+0.084$ correction factor discussed in Section~5, so our
final value, with 1$\sigma$ uncertainty, is $F_{bias}=0.24\pm 0.22$.
In Figure~12(d), the FBST relation from Figure~7(b) is
reproduced, and the locations of $\tau$~Boo~A and HD~189733A
are also shown.  Both exoplanet host stars lie slightly above
the main sequence star relation.  Extremely active stars
like EK~Dra, AB~Dor~A, and AU~Mic also lie above the relation
(see Figure~7), but neither $\tau$~Boo~A nor HD~189733A are in
this high-activity regime, with X-ray luminosities of
$\log L_X=28.76$ and $\log L_X=28.46$ \citep{js04},
respectively.  Other stars with these luminosities are
consistent with the FBST relation.  Thus, we do not believe
that high activity accounts for the anomalous coronal abundances
of $\tau$~Boo~A nor HD~189733A.  Considering that previous
observations have also suggested that stellar activity on these
stars is being affected by their exoplanets, it seems reasonable
to propose that the anomalously high $F_{bias}$ values are
also somehow caused by the exoplanet influence.

\section{Summary}

     We have conducted a survey of all main sequence star
{\em Chandra}/LETGS spectra with sufficient S/N for detailed
analysis.  This involves the analysis of 21
spectra from 19 stars, where both low-state and high-state
spectra are considered for $\alpha$~Cen~A and B.  From these
spectra we measure fluxes or upper limits for
118 lines from 49 ionic species.  The EM analyses based on these
measurements provide coronal temperature distributions and
abundance measurements, leading to the following conclusions:
\begin{enumerate}
\item In contrast to past analyses, we no longer find narrow
  EM peaks at $\log T=6.6$, though there is still generally an
  EM maximum near that temperature, and we also find a systematic increase
  of $+0.084$ in the ``FIP bias'' parameter, $F_{bias}$, compared
  to past measurements.  This is due to changes in Fe~XVII line
  emissivities in version~7.1 of the CHIANTI database, used here
  as the source for atomic data.  In particular, emissivities of
  three of the four strong Fe~XVII lines at $15-17$~\AA\ increased
  by $\sim 60$\% compared to past CHIANTI versions.
\item We expand on previous studies of the FBST relation,
  supplementing our LETGS sample with other published results from XMM
  and HETGS.  Consideration of Altair (A7~V) and $\eta$~Lep (F1~V)
  allows us to extend the relation to earlier spectral types than
  before.  We find that the relation is flat at A7-G5 spectral types,
  before increasing towards later types.  Replacing spectral type with
  $T_{eff}$ allows us to derive a quantitative FBST relation, which is
  provided in equation~(5).
\item Absolute coronal Fe abundances are quantified using a
  line-to-continuum analysis, focusing on the $25-40$~\AA\
  region.  We find a roughly linear correlation between
  [Fe/Fe$_*$] and $T_{eff}$,
  ${\rm [Fe/Fe_*]}=-1.288+2.176\times 10^{-4}~T_{eff}$, albeit
  with a lot of scatter.
\item While the solar and stellar coronal abundances are perfectly
  consistent with each other with regards to relative abundances as
  quantified by $F_{bias}$, such is not the case for absolute
  abundances.  While the solar FIP effect is generally accepted to
  involve low-FIP elements being enhanced, the stellar abundance
  measurements for solar-like G dwarfs seem more consistent with
  high-FIP depletions.  This discrepancy casts doubt on the
  reliability of the line-to-continuum analysis.
\item We measure Ne/O=0.39 for our sample of stars, consistent
  with previous work \citep{jjd05}.  We find no
  indication of activity dependence that might mitigate the
  discrepancy with the solar value, which is nevertheless less
  dramatic than it used to be if the larger solar measurement from
  \citet{pry18}, ${\rm Ne/O}=0.24\pm 0.05$, is utilized.
\item Coronal Si/Fe ratios are systematically high for M dwarfs.
  This may be due to the incorporation of Si into SiO in M dwarf
  photospheres.  We propose that the robustness of SiO leads
  to Si behaving more like a high-FIP element with regards to its
  coronal abundance.
\item Comparisons are made with the solar full-disk EM distributions
  from SDO/EVE \citep{sjs17}.  The corona of $\alpha$~Cen~B
  is the most solar-like in our sample, with a similar EM distribution
  and X-ray surface flux.
\item We find no change in coronal abundances between the low-state
  and high-state spectra of $\alpha$~Cen~A and B, suggesting that
  coronal abundances do not vary with activity cycles.
\item Our sample of main sequence stars provides a consistent picture
  for how EM distributions change with increasing $F_X$, with no
  evidence for any spectral type dependence.  Thus, we derive EM
  distributions as a function of $F_X$ (see Figure~10 and Table~7),
  which can be used as the basis for future theoretical studies of
  coronal heating, and can also be used to estimate EM distributions
  for any star with a measured broadband X-ray luminosity.
\item The stellar EM distributions are compared with solar QS, AR, and
  FL distributions.  The QS distribution, with $\log F_X=3.61$, may
  represent a minimum emission level for main sequence star coronae.
  Between $\log F_X=5.0$ and $\log F_X=6.5$, the stellar distributions
  agree very well with the solar AR distribution for $\log T<6.6$, in
  terms of both magnitude and slope, consistent with the idea that the
  surfaces of moderately active stars are completely filled with
  solar-like ARs.  The sum of the AR and FL distributions represents a
  reasonable approximation of the stellar distribution we find at
  $\log F_X\sim 6.0$.  Completely covering the surface of a star with
  the X-class flares represented by our FL distribution leads to a
  decent approximation of the observed EM distribution of our most
  active stars, with $\log F_X\sim 7.5$.
\item In deriving the solar FL distribution, we also inferred a relation
  between GOES X-class and X-ray luminosity: $\log L_X=0.56\log X_f+27.75$.
\item The coronal abundances of the exoplanet host $\tau$~Boo~A are
  anomalous.  The $F_{bias}$ value is somewhat higher than it
  should be, and [Fe/Fe$_*$] is much lower.  An $F_{bias}$ estimate
  for another exoplanet host, HD~189733A, from XMM measurements
  is also somewhat higher than it should be.  We therefore conclude
  that the close-in, massive exoplanets of these stars may be
  affecting their coronal abundances.
\end{enumerate}


\acknowledgments

Support for this work was provided by NASA through Chandra Award
Numbers GO6-17002Z, AR7-18003Z, and GO8-19002Z issued by the Chandra
X-ray Center (CXC), which is operated by the Smithsonian Astrophysical
Observatory for and on behalf of the National Aeronautics Space
Administration under contract NAS8-03060.  This research has made use
of the SIMBAD database, operated at CDS, Strasbourg, France.

\software{CIAO \citep[v4.9;][]{af06}, PINTofALE \citep[v2.97;][]{vk00}}

\end{document}